\documentclass[prb,twocolumn,showpacs,amsmath,amssymb,superscriptaddress]{revtex4-2}
\usepackage{epsfig,epstopdf} 
\usepackage{graphicx}
\usepackage{dcolumn}
\usepackage{bm}
\usepackage{bbm}
\usepackage{ulem}
\usepackage{xcolor}
\usepackage{amsmath}
\usepackage{amsthm}
\usepackage{esint}
\usepackage{bbold}
\usepackage{slashed}
\usepackage{hyperref}
\usepackage{mathrsfs}
\usepackage{subfigure}
\usepackage{verbatim}
\usepackage{dsfont}
\usepackage{float}

\newcommand{\nc}{\newcommand}
\nc{\be}{\begin{equation}} \nc{\ee}{\end{equation}}
\nc{\bea}{\begin{eqnarray}} \nc{\eea}{\end{eqnarray}}
\nc{\bean}{\begin{eqnarray*}} \nc{\eean}{\end{eqnarray*}}
\nc{\dg}{\dagger}
\nc{\ua}{\uparrow} \nc{\da}{\downarrow}
\nc{\lag}{\langle} \nc{\rag}{\rangle}

\begin{document}

\bibliographystyle{apsrev4-1}

\title{Kitaev Building-block Construction for Inversion-Protected Higher-order Topological Superconductors}
\author{Rui-Xing Zhang}
\email{ruixing@utk.edu}
\affiliation{Department of Physics and Astronomy, The University of Tennessee, Knoxville, TN 37996, USA}
\affiliation{Department of Materials Science and Engineering, The University of Tennessee, Knoxville, TN 37996, USA}
\affiliation{Institute of Advanced Materials and Manufacturing, The University of Tennessee, Knoxville, TN 37920, USA}
\affiliation{Condensed Matter Theory Center and Joint Quantum Institute, Department of Physics, University of Maryland, College Park, Maryland 20742-4111, USA}
\author{Jay D. Sau}
\email{jaydsau@umd.edu}
\affiliation{Condensed Matter Theory Center and Joint Quantum Institute, Department of Physics, University of Maryland, College Park, Maryland 20742-4111, USA}
\author{S. Das Sarma}
\affiliation{Condensed Matter Theory Center and Joint Quantum Institute, Department of Physics, University of Maryland, College Park, Maryland 20742-4111, USA}

\begin{abstract}
We propose a general theoretical framework for both constructing and diagnosing inversion-protected higher-order topological superconductors using Kitaev building blocks, a higher-dimensional generalization of Kitaev's one-dimensional Majorana model. For a given crystalline symmetry, the Kitaev building blocks serve as a complete basis to construct all possible Kitaev superconductors that satisfy the symmetry requirements. We derive a simple yet powerful Majorana counting rule that can unambiguously diagnose the existence of higher-order topology for all Kitaev superconductors. We expect this real-space diagnosis to work for general two-dimensional higher-order topological superconductors within this symmetry class. As proof of concept, we have identified two inequivalent stacking strategies using the Kitaev building blocks, based on which we have constructed minimal tight-binding models with symmetry-protecetd Majorana corner modes. Moreover, we have successfully applied our diagnosis to comprehend the Majorana corner physcis in a superconductor model with a fragile Wannier obstruction, confirming the validity of our theory beyond the Kitaev limit. Our work paves the way for interpreting higher-order topological superconductivity from the real-space perspective.
\end{abstract}

\date{\today}
\maketitle

\section{Introduction}

The concept of topology has revolutionized our understanding of condensed matter systems in the past decades. The revolution started with the quantum Hall effect \cite{laughlin1983,thouless1982,avron1983} and continued through the seminal works on the 10-fold way \cite{zirnbauer1996,altland1997} and topological insulators (TI) \cite{kane2005z2}, becoming a dominant theme in condensed matter physics over the last 10 years, leading to the exciting concept of topological quantum computation using non-Abelian anyons \cite{kitaev2001,nayak2008}. Even within the same symmetry class, there could exist several types of topologically distinct phases that cannot be connected through an adiabatic evolution path and thus behave differently in various aspects. In particular, the topological properties for a large class of systems are only well-defined when certain types of symmetries are present. This class of topological systems is known as the symmetry-protected topological (SPT) state \cite{chen2013symmetry,senthil2015spt,chiu2016classification}, and most, if not all, currently known free-fermion topological states are technically SPT phases even if this is not always explicitly mentioned. When placed on an open geometry, the bulk topology of a $D$-dimensional SPT system enforces the existence of anomalous in-gap modes on its $(D-1)$-dimensional boundary, which cannot be removed without either closing the bulk energy gap or breaking the protection symmetry. Such in-gap boundary modes are ``anomalous" in the sense that they can never be realized in any $(D-1)$-dimensional bulk system -- they are strictly the boundary modes corresponding to the bulk topology, an example of a bulk-boundary correspondence.

For free-fermion SPT systems such as topological insulators, topological band theories are extremely successful in classifying and predicting new topological materials \cite{hasan2010colloquium,qi2011topological}. To capture various band topology, one direct approach is to mathematically define the corresponding topological invariants for band insulators with different internal or crystalline symmetries \cite{fu2011topological,hsieh2012topological,slager2013space,liu2014topological,ando2015topological,zhang2015topological,fang2015new,wang2016hourglass,chang2017mobius,chiu2016classification}. On the other hand, all known topological band insulators present obstruction to a symmetric and localized Wannier function description \cite{thonhauser2006insulator,soluyanov2011wannier}. Therefore, Wannierizability can be treated as a diagnosis for distinguishing topological and trivial band insulators. Notably, the recent breakthrough in the band representation theory provides us with a complete list of all possible trivial atomic insulators for all space groups \cite{bradlyn2017TQC}. Consequently, topological systems can be systematically sorted by simply excluding the known atomic limits for band insulators. This is, in principle, a conceptually revolutionary new way of classifying insulators, connecting quantum chemistry (i.e. the atomic limit) with solid state band theories.

It was recently realized that some topological insulators protected by lattice symmetries admit a higher-order version of the bulk-boundary correspondence \cite{highertopo}, which are dubbed higher-order topological insulators \cite{benalcazar2017electric,benalcazar2017quantized,langbehn2017reflection,song2017d,schindler2018higher, khalaf2018higher,miert2018higher,you2018higher,wang2019higher,you2020higher}. Specifically, the $D$-dimensional bulk topology in these systems is indicated by anomalous in-gap modes on their $(D-n)$-dimensional boundary with $n\in\{2,...,D-1\}$. Some three-dimensional (3d) axion insulator candidates \cite{varnava2018surface,wieder2018axion}, including EuIn$_2$As$_2$ \cite{xu2019higher}, Bi$_{2-x}$Sm$_x$Se3 \cite{yue2019symmetry}, and MnBi$_{2n}$Te$_{3n+1}$ \cite{zhang2019mobius}, have been theoretically proposed to host 1d inversion-protected chiral fermion channels that live on the ``hinges" connecting two neighboring gapped surfaces. Meanwhile, experimental signatures of 1d inversion-protected helical hinge modes have been observed in Bismuth using scanning tunneling microscopy \cite{schindler2018Bi}. Similar to a conventional TI, a higher-order TI is necessarily Wannier obstructed. As a result, by filtering out Wannierizable atomic insulators, researchers have designed/derived simple functions of crystalline symmetry eigenvalues at high-symmetry momenta, which can correctly diagnose the higher-order topology. These symmetry-eigenvalue-based functions are known as ``symmetry indicators" \cite{kruthoff2017topological,po2017symmetry,khalaf2018symmetry,ono2018unified}.  

In this work, our main focus is the higher-order version of topological superconductors (TSC), which has been recently under active research in the community \cite{wang2018high,yan2018majorana,liu2018majorana,wang2018weak,shapourian2018topological,hsu2018majorana,bultinck2019three,zhang2019helical,zhang2019higher,wu2022high,hsu2019inversion,zhang2019dirac,yan2019higher}. In particular, a 2d higher-order TSC is featured by zero-dimensional (0d) corner-localized Majorana zero modes, which potentially offer new promising platforms for Majorana-based topological quantum computation \cite{pahomi2019braiding,zhang2020topological}. Analogous to the theoretical developments of crystalline TIs, theories of symmetry indicators in TSCs have been developed by multiple groups, which have been proved quite successful in characterizing the topological nature of various crystalline superconducting phases \cite{benalcazar2014classification,ono2019symmetry,shiozaki2019variants,skurativska2020atomic,ono2019refined,geier2019symmetry,roberts2019second}. Despite many similarities, however, we note that topological descriptions for insulators and superconductors also feature several conceptual distinctions.

First of all, the physical meaning of being ``atomic" for a superconductor is unclear. As is known, the most important length scale for superconductors is the superconducting coherence length $\xi$, which is about $10^1\sim 10^3$ atomic lattice constants. This directly implies that the atomic-scale microscopic physics should not be crucial for describing superconductivity and its related topological phenomena. In other words, any topology-enforced Majorana physics for a TSC should just arise from the low-energy physics near the Fermi surface, where the energy cut-off is around $\xi^{-1}$ and not a microscopic atomic lattice length scale. This feature clearly distinguishes between the physical meanings of ``atomic limit" for superconductors and insulators. 

Second, while the triviality of atomic band insulators can be diagnosed from its Wannierizability, the conventional wisdom indicates that {\it a Wannierizable superconductor can also be topological}. As perhaps the simplest example, the 1d Kitaev Majorana chain \cite{kitaev2001} is not only Wannierizable but also hosting 0d Majorana end modes. Clearly, the 0d Majorana modes are anomalous boundary modes and cannot be realized in any 0d electron systems. Therefore, it is natural to speculate that a 2d higher-order topological superconductor with 0d Majorana corner modes could also be Wannierizable, similar to a Kitaev chain \cite{kitaev2001}.

Motivated by the above considerations, we propose in this work the concept of {\it Kitaev limit} as a direct higher-dimensional generalization for Kitaev's 1d Majorana chain, which offers a natural language for describing higher-order TSCs. A BdG system satisfying the Kitaev limit is termed a ``Kitaev superconductor" which is always gapped and Wannier representable. In particular, each ground-state Wannier orbital of a Kitaev superconductor is always formed by binding only a pair of Majorana fermions. Therefore, the notion of Kitaev limit should be viewed as a ``minimalism" decription of Wannierizability, which greatly benefits both topology characterizations and model constructions.

Focusing on 2d class D systems with a spatial inversion symmetry, we identify four inequivalent minimal Kitaev superconductors (the Kitaev building blocks) as a complete set of basis for building general Kitaev superconductors. In particular, by stacking these Kitaev building blocks, we are able to systematically construct various Kitaev superconductors with intrinsic higher-order topology. The building-block picture motivates us to propose a simple but powerful Majorana counting rule as a set of real-space topological indices to diagnose the existence of symmetry-protected Majorana corner modes.  To confirm our Majorana counting rule, we present two different stacking strategies with the Kitaev building blocks that have in principle exhausted all possible configurations for Kitaev superconductors. For each stacking strategy, we derive explicit ``stacking recipes" based on our counting rule, which efficiently predict possible stacking configurations with higher-order topology. To demonstrate these ideas, we provide several minimal models for distinct stacking strategies and confirm their higher-order topological nature. 

The great success of the Majorana counting rule further suggests that its applicability could go beyond not only the 2d Kitaev limit, but also the constraint of being Wannierizable. As a proof of concept, we consider a higher-order TSC with a fragile Wannier obstruction, which can be decomposed into a linear superposition of Kitaev limits. By performing the counting rule for each constituent block, we have successfully explained the corner-mode physics in this system. Note that stably Wannier-obstructed systems are usually first-order topological in 2D, which is beyond the scope of our work.     

The introduction of the ``Kitaev limit" as the fundamental building blocks for higher-order topological superconductors as well as the introduction of the new Majorana counting rule are the two important new concepts in our work. We show that these new concepts enable both the construction of higher-dimensional topological superconductors as well as the diagnosis for already-proposed higher-order topological superconductors, thus establishing our ideas as the foundation for the ``Quantum Chemistry" of Topological Superconductors.     

This paper is organized as follows. In Sec. \ref{Sec: Main Results}, we first present the main results of our work providing our definitions of higher-order topology, the Kitaev limit, and the Majorana counting rule. In Sec. \ref{Sec: Classification}, we define the Kitaev building blocks and classify all Kitaev superconductors with the concept of 2d polarization. We then explicitly derive the Majorana counting rule as a necessary and sufficient condition for higher-order topology in Kitaev superconductors. In Sec. \ref{Sec: Examples}, we provide explicit constructions of Kitaev superconductors with higher-order topology to demonstrate our Majorana counting rule for different stacking strategies. In Sec. \ref{Sec: Conjecture}, we propose a real-space diagnosis for all 2d class-D higher-order TSCs with an inversion symmetry. This conjecture is verified in Sec. \ref{Sec: Fragile} for a realistic example that hosts both higher-order topology and fragile Wannier obstruction. The conclusion and discussions about future directions are presented in Sec. \ref{Sec: Conclusion}.

\section{Main Results}\label{Sec: Main Results}

This section summarizes our main results in this work. We first provide a definition of symmetry-protected higher-order topology for 2d class D superconductors. We then define the Kitaev limit, a key concept in this work and a natural language for describing higher-order topological superconductors. Finally, we present a simple Majorana counting rule as a higher-order topological diagnosis for general Kitaev superconductors.       

\subsection{What is a Higher-order Topological Superconductor?}

A 2d class D superconductor is defined to have symmetry-protected higher-order topology if 
\begin{enumerate}
	\item[(i)] its bulk BdG spectrum is gapped; 
	\item[(ii)] it has no 1d anomalous Majorana boundary modes; 
	\item[(iii)] it has 0d Majorana zero modes exponentially localized on its symmetric boundary;
	\item[(iv)] while preserving the symmetry, the Majorana zero modes cannot be eliminated without closing the bulk gap.
\end{enumerate}

This definition is inspired by the phenomenological fact that a 2d higher-order TSC generally features gapped edge states and 0d boundary Majorana modes that are most likely localized around the corners. Notably, the presence of 0d boundary Majorana modes does not guarantee a bulk higher-order topology by itself. It is the necessary crystalline-symmetry protection that promotes such 0d boundary Majorana modes to a hallmark of nontrivial bulk topology.

The crucial role of symmetry protection can be understood as follows. Consider a 2d higher-order TSC with a (crystalline) symmetry ${\cal G}$ on a symmetry-preserving open geometry. We then define ${\cal N_G}$ as the number of 0d Majorana modes on the 1d edge ``modulo" the symmetry ${\cal G}$. Namely, if two distinct boundary Majorana modes are related by ${\cal G}$, they will be counted as the same Majorana mode in defining ${\cal N_G}$. If ${\cal N_G}$ is odd, it is obvious that the 0d Majorana modes cannot be eliminated by any ${\cal G}$-preserving edge perturbation (e.g. attaching 1d Kitaev chains in a ${\cal G}$-symmetric way). This robustness of 0d Majorana modes is exactly a manifestation of ${\cal G}$-protected higher-order bulk-boundary correspondence. This analysis also provides an alternative yet equivalent definition that {\it a 2d class D superconductor is ${\cal G}$-protected higher-order topological if and only if ${\cal N_G}$ is odd}. \\          

\begin{figure}[t]
	\centering
	\includegraphics[width=0.45\textwidth]{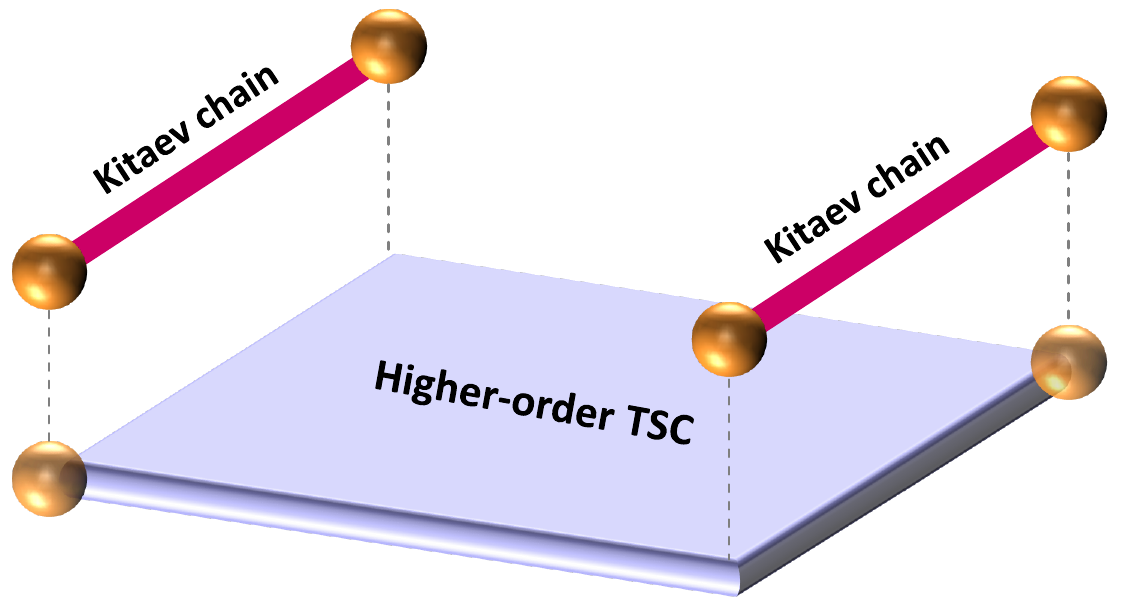}
	\caption{A schematic of 2d inversion-protected higher-order TSC with a pair of corner-localized Majorana modes. To preserve the inversion symmetry, the Kitaev chains attached to the edge must come in pairs. Since a pair of Kitaev chains will necessarily add 4 Majorana zero modes to the edge, the corner Majorana physics cannot be essentially eliminated.}
	\label{Fig: higher-order schematic}
\end{figure}

In this work, we focus on the BdG systems with 2d spatial inversion symmetry ${\cal I}$ to demonstrate our general theory of higher-order TSCs. Practically, ${\cal N_I}$, the number of Majorana zero modes modulo ${\cal I}$, can be simply counted on one $x$-edge and one $y$-edge, as well as the corner shared by the edges. Then by definition, an inversion-protected higher-order TSC is typically featured by two Majorana modes that are spatially separated at the opposite corners \cite{khalaf2018higher}. As a schematic example in Fig. \ref{Fig: higher-order schematic}, symmetrically coupling Kitaev chains to the boundary of an inversion-symmetric higher-order TSC can only shift the position of corner Majorana modes, but can never eliminate them. Therefore, this boundary configuration is stable against any boundary perturbations that preserve inversion symmetry, a hallmark of inversion-protected higher-order topology.

The higher-order TSCs that will be discussed in this work are Wannierizable, similar to recently proposed electronic insulators with fractional corner charges \cite{benalcazar2017quantized,benalcazar2019quantization,schindler2019fractional,zhu2019identifying}. Notice that the Wannierizability of corner-charged electron systems prevent them from having stable gapless bound states at the edges, since such bound state can always be removed by an applied smooth potential at the edge. In contrast, in-gap 0d Majorana modes on the boundary of a higher-order topological superconductor are robustly pinned at zero energy by the particle-hole symmetry. This crucial difference thus allows the existence of Wannierizable topological superconductors, which turn out to be most likely higher-order topological. This is one important motivation for us to consider Wannierizable higher-order topological BdG systems in this work.      

\subsection{Kitaev Limit for Superconductors} \label{Sec: Kitaev limit}

Higher-order topological superconductors are defined by the existence of anomalous unpaired Majorana modes at the boundaries. Following Kitaev \cite{kitaev2001}, representing a BdG Hamiltonian in terms of Majorana fermions provides a natural representation to include the particle-hole symmetry. However, such Majorana representation introduces an additional constraint, which has no analog in the case of insulators, that Majorana fermions must appear in pairs on any so-called atomic site. For the purpose of superconductors, an "atomic site" ${\bf R}$ is any position that we place a pair of Majorana fermions (or equivalently an electron and a hole) in the discretized BdG Hamiltonian. The on-site transformation between an electron operator $c_R^\dagger$ and the corresponding pair of Majorana fermions $\alpha_{\bf R}$ and $\beta_{\bf R}$ is given by $c^{\dagger}_{\bf R}=(\alpha_{\bf R} - i \beta_{\bf R})/\sqrt{2}$.

\begin{figure}[t]
	\centering
	\includegraphics[width=0.49\textwidth]{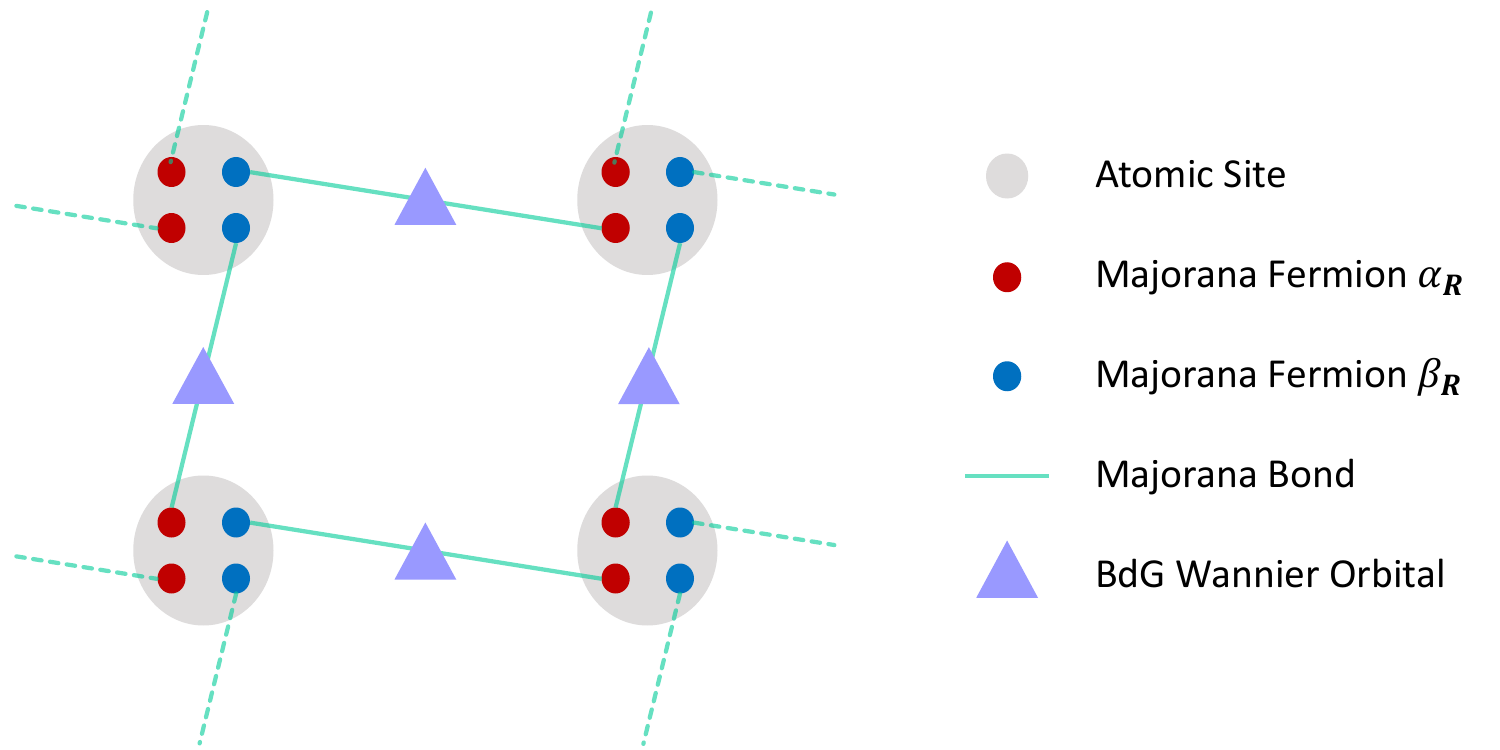}
	\caption{An example of Kitaev limit with two pairs of Majorana fermions within each unit cell. Each Majorana fermion has exactly one Majorana bond attached. The maximally localized Wannier orbital sits at each bond center.}
	\label{Fig: Kitaev Limit}
\end{figure}

A big advantage of the Majorana representation is that the electron-electron, hole-hole, and electron-hole couplings now become coupling or bondings among the Majorana operators. As shown in Fig. \ref{Fig: Kitaev Limit}, when two Majorana fermions (the red and blue dots) are coupled with each other, we can pictorially connect them with a line to demonstrate the existence of a Majorana bond. In this work, we only consider the Majorana bonds that are local in space. Based on the Majorana representation, we define a special class of BdG systems:

\begin{itemize}
	\item In the Majorana representation, a BdG system with periodic boundary conditions is in the Kitaev limit, if every Majorana fermion is attached to {\it exactly }one Majorana bond. Such BdG system is termed as a Kitaev superconductor.
\end{itemize}

The Kitaev limit is one of the key concepts in our work. It is physically motivated by the pioneering work of Kitaev \cite{kitaev2001}, in which he found a special limit of a 1d spinless p-wave superconductor model (often known as a Kitaev chain) that is exactly solvable. While Kitaev's original idea aims at understanding the boundary physics in 1d class D TSCs, we generalize this concept to two and higher dimensions for any symmetry class \cite{kitaevlimit}. We will see that the concept of Kitaev limit provides a natural framework to describe higher-order topological superconductors with 0d Majorana zero modes.

We emphasize that all Kitaev superconductors are featured by both a {\it bulk energy gap} and a description with {\it maximally localized BdG Wannier functions}. The bulk gap arises from the fact that every Majorana fermion is uniquely paired with another nearby Majorana fermion. The Majorana bond also leads to a exponentially localized Wannier function sitting at the bond center. For a Kitaev superconductor with bonds connecting only on-site or nearest-neighboring Majorana fermions, the range of the Wannier function is restricted within one unit cell. Adding any additional Majorana bonds to this Kitaev superconductor will only further delocalize the spatial profile of the Wannier function. This is why the Wannier functions of a Kitaev superconductor are maximally localized. Since the Majorana representation and the Wannier representation with maximally localized Wannier orbitals are essentially the two sides of the same coin, they will be used interchangeably in this work.   

A schematic example of a 2d Kitaev superconductor is shown in Fig. \ref{Fig: Kitaev Limit}. Within one unit cell, we have considered two atomic sites (the gray disks) that coincide with each other \cite{site}. Each atomic site hosts a pair of Majorana fermions $\alpha_{\bf R}$ (the red dot) and $\beta_{\bf R}$ (the blue dot). In particular, every Majorana fermion is only connected to one distinct Majorana fermion through a intersite Majorana bond (the green line), which clearly satisfies the definition of Kitaev limit. For demonstration, we also plot the maximally localized BdG Wannier orbitals (the purple triangles) sitting at the bond center.

Our definition of the Kitaev limit relies on physical Majorana degrees of freedom and thus distinguishes a superconducting BdG system from a particle-hole symmetric electron system with no superconductivity. This fact is crucial for discussing a topological classification for superconductors. 

\subsection{Condition of Higher-order Topology for Kitaev Superconductors} \label{Sec: Majorana counting rule}

If an inversion-symmetric Kitaev superconductor has $n_i^{(A)}$ atomic sites and $n_i^{(W)}$ Wannier orbitals at maximal Wyckoff position ${\bf q}_{1i}$ for $i\in\{a,b,c,d\}$, it has inversion-protected higher-order topology {\it if and only if}
\begin{equation}
\Delta_{b,c,d} \equiv 1 \ \ (\text{mod }2).
\end{equation}    
where we have defined a set of Majorana counting numbers as
\begin{equation}
	\Delta_i = n_i^{(W)} - n_i^{(A)}. 
\end{equation}
This simple counting of bulk atomic sites and Wannier orbitals provide a necessary and sufficient condition for a general Kitaev superconductor to be higher-order topological, which is thus dubbed the ``Majorana counting rule". 

For 2d inversion-symmetric systems, we have defined the maximal Wyckoff positions as the real-space positions invariant under ${\cal I}$ operation, up to lattice translations. Within one unit cell, we label the maximal Wyckoff positions as 
\begin{eqnarray}
{\bf q}_{1a} &=& (0,0),\ {\bf q}_{1b} = (\frac{1}{2},0),\nonumber \\
{\bf q}_{1c} &=& (0,\frac{1}{2}),\ {\bf q}_{1d} = (\frac{1}{2},\frac{1}{2}), 
\end{eqnarray} 
where the lattice constants are set to be ${\bf a}_x = {\bf a}_y=1$ for simplicity.

Physically, the position information of atomic sites determines how crystalline symmetries are implemented in a BdG system, which provides necessary symmetry constraint for choosing a symmetric open boundary. Therefore, it is the countings of both atomic sites and Wannier orbitals that together determine the boundary Majorana information on a symmetric Kitaev superconductor.    

We will explicitly derive the Majorana counting rule in the coming Sec. \ref{Sec: Deriving Counting Rule}. The concept of Kitaev limit and the Majorana counting rule are the most important results in this work.

\section{Classification of Kitaev Superconductors and Majorana Counting} \label{Sec: Classification}

In this section, we will elaborate on our main results summarized in the previous section by systematically classifying 2d Kitaev superconductors and explicitly deriving the Majorana counting rule as a powerful diagnosis for higher-order topology.

\subsection{Kitaev Building Blocks}

\begin{figure*}[t]
	\centering
	\includegraphics[width=0.99\textwidth]{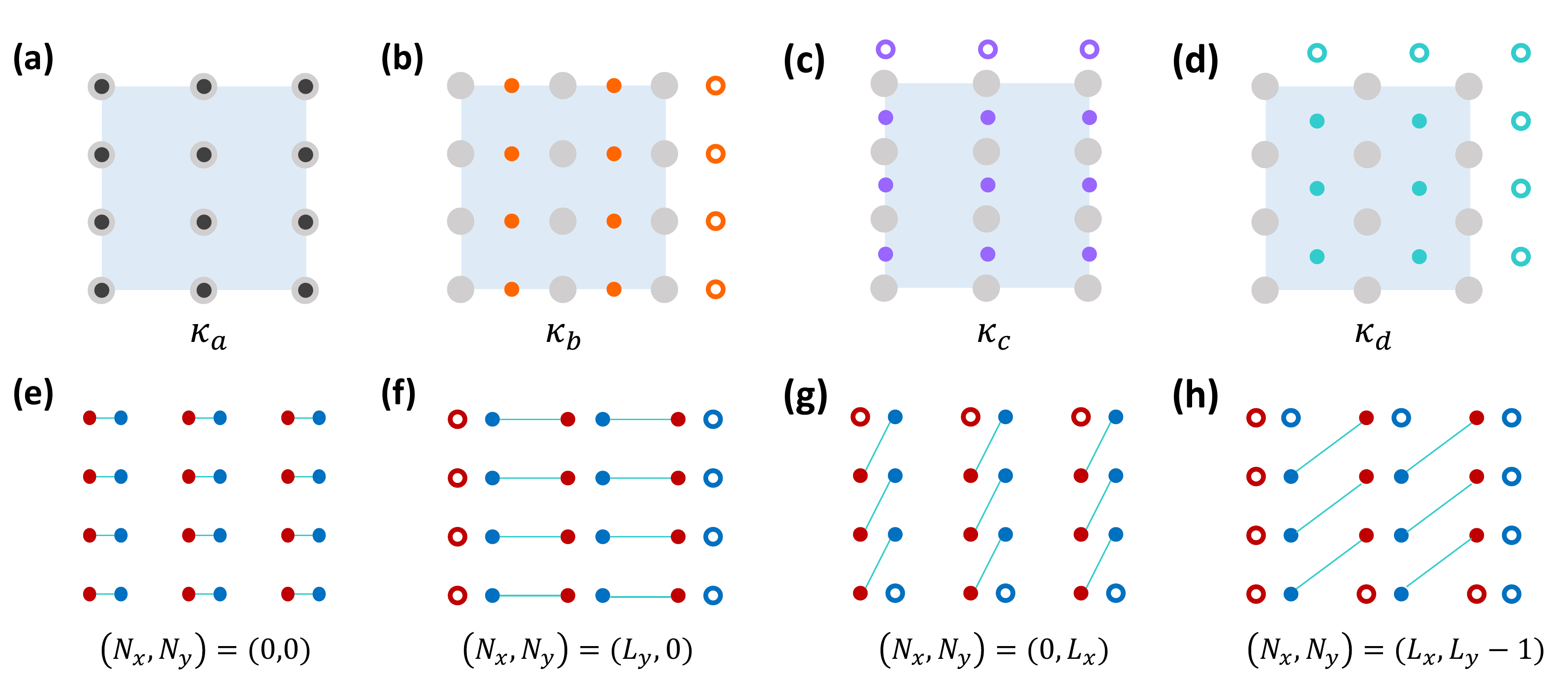}
	\caption{The Wannier representations and the Majorana representations for each Kitaev building block are shown in (a) - (d) and (e) - (h), respectively. The dangling Wannier orbitals and Majorana zero modes are shown in colored circles. $(N_x,N_y)$ shows the numbers of dangling Majorana zero modes for one $x$ edge and one $y$ edge, respectively. }
	\label{Fig: Atomic Polarization}
\end{figure*}

The key to classify all these Kitaev superconductors for a specific symmetry class is to identify a set of minimal/simplest symmetry-allowed Kitaev superconductors as a complete basis, based on which one can construct more complicated situatons. This set of minimal Kitaev superconductors are termed ``Kitaev building blocks". For a symmetry group $G$, the $d$-dimensional Kitaev building blocks are defined as all $d$-dimensional Kitaev superconductors that (i) preserve all symmetries in $G$; (ii) can NOT be decomposed into a superposition of any other $G$-preserving $d$-dimensional Kitaev superconductors.

For 2d class D Kitaev superconductors with inversion symmetry, there exist four distinct Kitaev building blocks $\kappa_i$ with $i\in\{a,b,c,d\}$, which are constructed by placing one atomic site at ${\bf q}_{1a}$ and one {\it maximally localized} BdG Wannier orbital at a maximal Wyckoff position ${\bf q}_{1i}$ in the Wannier representation.

While a Wannier orbital could be even (s-like) or odd (p-like) under inversion operation, the specific orbital type is irrelevant for our most discussions. In Fig. \ref{Fig: Atomic Polarization} (a) - (d), we schematically plot those four Kitaev building blocks $\kappa_{a,b,c,d}$. As before, we use the gray disk at ${\bf q}_{1a}$ to denote the atomic sites and the colored dots to denote the maximally localized Wannier orbitals. Map the Wannier representation back to the Majorana representation, we arrive at Fig. \ref{Fig: Atomic Polarization} (e) - (h), which demonstrate the Majorana bonding configurations for the Kitaev building blocks.

In principle, one can always place a Wannier orbital on some non-maximal Wyckoff position ${\bf q}_{\alpha}$ that is not invariant under inversion operation ${\cal I}$. Then we are required to place another Wannier orbital at a different but inversion-related Wyckoff position ${\cal I}{\bf q}_{\alpha}$ just to preserve the inversion symmetry. On the other hand, we can always simultaneously move the Wannier orbital at ${\bf q}_{\alpha}$ and the other one at ${\cal I}{\bf q}_{\alpha}$ to any maximal Wyckoff position in an adiabatic and symmetric way. Therefore, putting Wannier orbitals on non-maximal Wyckoff positions is always equivalent to a double-stacking of some Kitaev building blocks $\kappa_i$. This is why the four Kitaev building blocks in Fig. \ref{Fig: Atomic Polarization} form a complete basis set for general Kitaev superconductors.

\subsection{Polarization and Topological Classification for Kitaev Superconductors}

To further characterize the topological properties of Kitaev building blocks, we define a 2d polarizaton ${\cal P}=(P_x, P_y)$ for Kitaev superconductors as the net relative displacement vector between the atomic sites and the Wannier orbitals, where $P_{x,y}$ are defined modulo the lattice constants. By definition, the polarization ${\cal P}_{\kappa_i}$ for a Kitaev building block $\kappa_{i}$ is exactly the Wyckoff position of its Wannier orbital,
\begin{equation}
	{\cal P}_{\kappa_i} = {\bf q}_{1i}.
\end{equation} 
It should be emphasized that our definition for the polarization ${\cal P}$ of Kitaev superconductors is {\it purely geometric}. In particular, ${\cal P}$ should NOT be confused with the physical charge polarization for insulators \cite{resta1994}, which is only well-defined in the presence of charge $U(1)$ symmetry. Here we use the term ``polarization" for ${\cal P}$ just to follow the convention in electron systems. 

It is straightforward to see that inversion symmetry requires both components of polarization vector $P_{x,y}$ to be quantized to either $0$ or $1/2$ modulo 1. In addition, we find the quantized value of ${\cal P}$ is directly linked with weak topological phenomena. When $P_{x(y)}=1/2$, the Kitaev superconductor displays weak topology \cite{seroussi2014topological}, with its edge normal to the $x(y)$-direction hosting a non-degenerate edge Majorana flat band. 

A pictorial understanding of the weak topology for the Kitaev building blocks is clearly explained in Fig. \ref{Fig: Atomic Polarization} (e) - (h). Take the building block $\kappa_d$ with ${\cal P}=(1/2, 1/2)$ as an example. As shown in Fig. \ref{Fig: Atomic Polarization} (h), $\kappa_d$ on an open geometry clearly hosts unpaired Majorana zero modes [colored circles in Fig. \ref{Fig: Atomic Polarization} (h)] in both $x$ and $y$ edges. If we calculate the edge spectrum for $\kappa_d$, the unpaired Majorana modes will form a single flat band at exactly zero energy for both $x$ and $y$ edges, a hallmark for 2d weak TSCs. Similarly shown in Fig. \ref{Fig: Atomic Polarization} (b) and (c), $\kappa_b$ and $\kappa_c$ also host non-degenerate edge Majorana flat bands on their $x$ and $y$ edges, respectively. Their ${\cal P}$ values agree with the existence of weak topology. Since all Kitaev superconductors can be constructed with the Kitaev building blocks, it is clear that this bulk-boundary correspondence between ${\cal P}$ and edge Majorana bands should generally hold.

For each building block with open boundary conditions, it is straightforward to count the dangling Majorana zero modes on the boundary. In Fig. \ref{Fig: Atomic Polarization} (e) - (h), we use $N_x$ and $N_y$ to respectively denote the numbers of dangling Majorana modes on each $x$ and $y$ edge. We explicitly show how $N_{x,y}$ scale with $L_{x,y}$, which are the numbers of unit cells along $x$ and $y$ directions. Notably, $(N_x, N_y)$ will provide an important input for deriving the Majorana counting rule in Sec. \ref{Sec: Deriving Counting Rule}. 

\begin{figure}[t]
	\centering
	\includegraphics[width=0.4\textwidth]{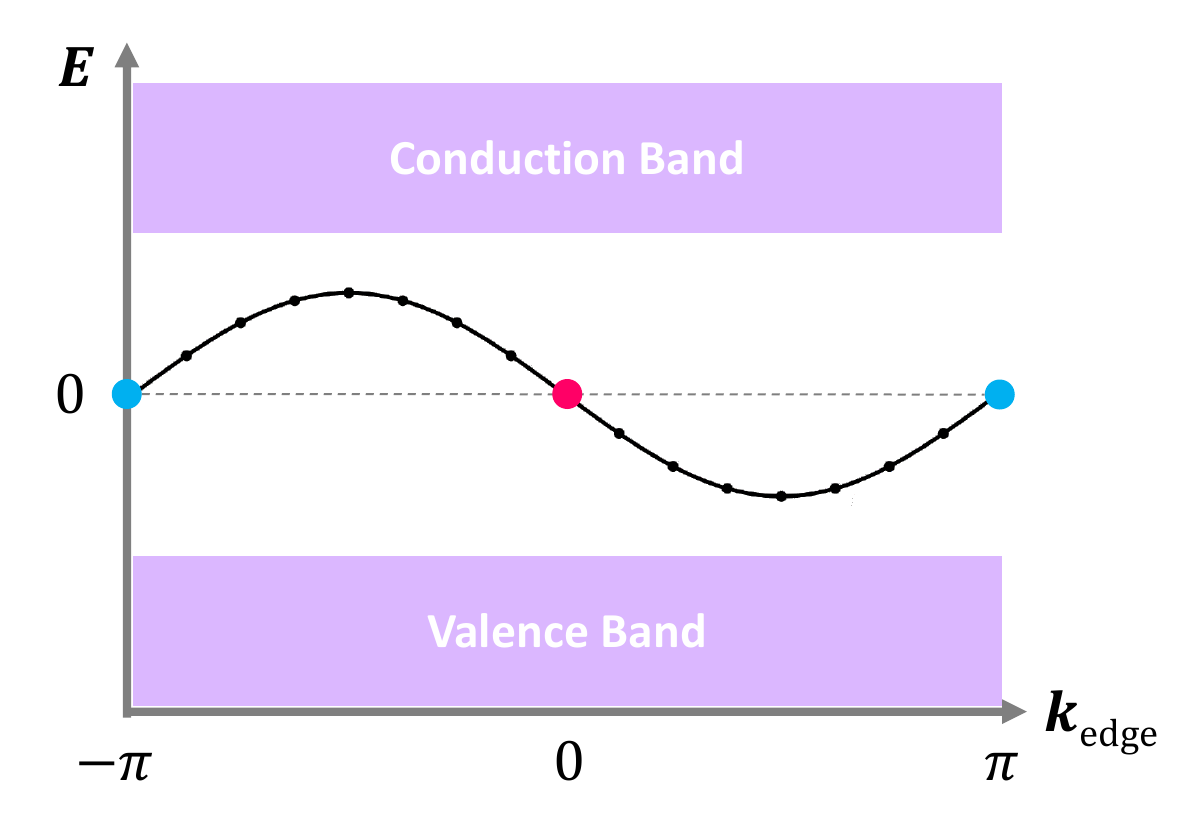}
	\caption{A schematic for a dispersing edge Majorana band in a weak TSC. This edge band is anomalous and is enforced to cross zero energy at high-symmetry momenta (colored dots) in the edge Brillouin zone.}
	\label{Fig: Edge Majorana Band}
\end{figure}

In principle, one can go beyond the Kitaev limit and deform a completely flat edge Majorana band into a dispersing one, as shown in Fig. \ref{Fig: Edge Majorana Band}. Such dispersing Majorana edge band is also anomalous and cannot be realized in any 1d bulk BdG system. This is because particle-hole symmetry will require the edge Majorana band to cross zero energy at $k_\text{edge} = 0, \pi$, manifesting its Majorana nature. 

With the bulk-boundary correspondence of ${\cal P}$, we now classify all 2d Kitaev superconductors into three topologically inequivalent classes: (i) trivial Kitaev superconductors with no weak or higher-order topology; (ii) weak TSCs with ${\cal P}\neq 0$; (iii) higher-order TSCs with ${\cal P}=0$.

 Even beyond the Kitaev limit, we still expect that this topological classification should generally hold for any 2d Wannierizable superconductors.  

\subsection{Atomic Site, Stacking, Boundary Conditions, and Symmetry Representation}\label{Sec: Atomic Sites}

\begin{figure}[t]
	\centering
	\includegraphics[width=0.47\textwidth]{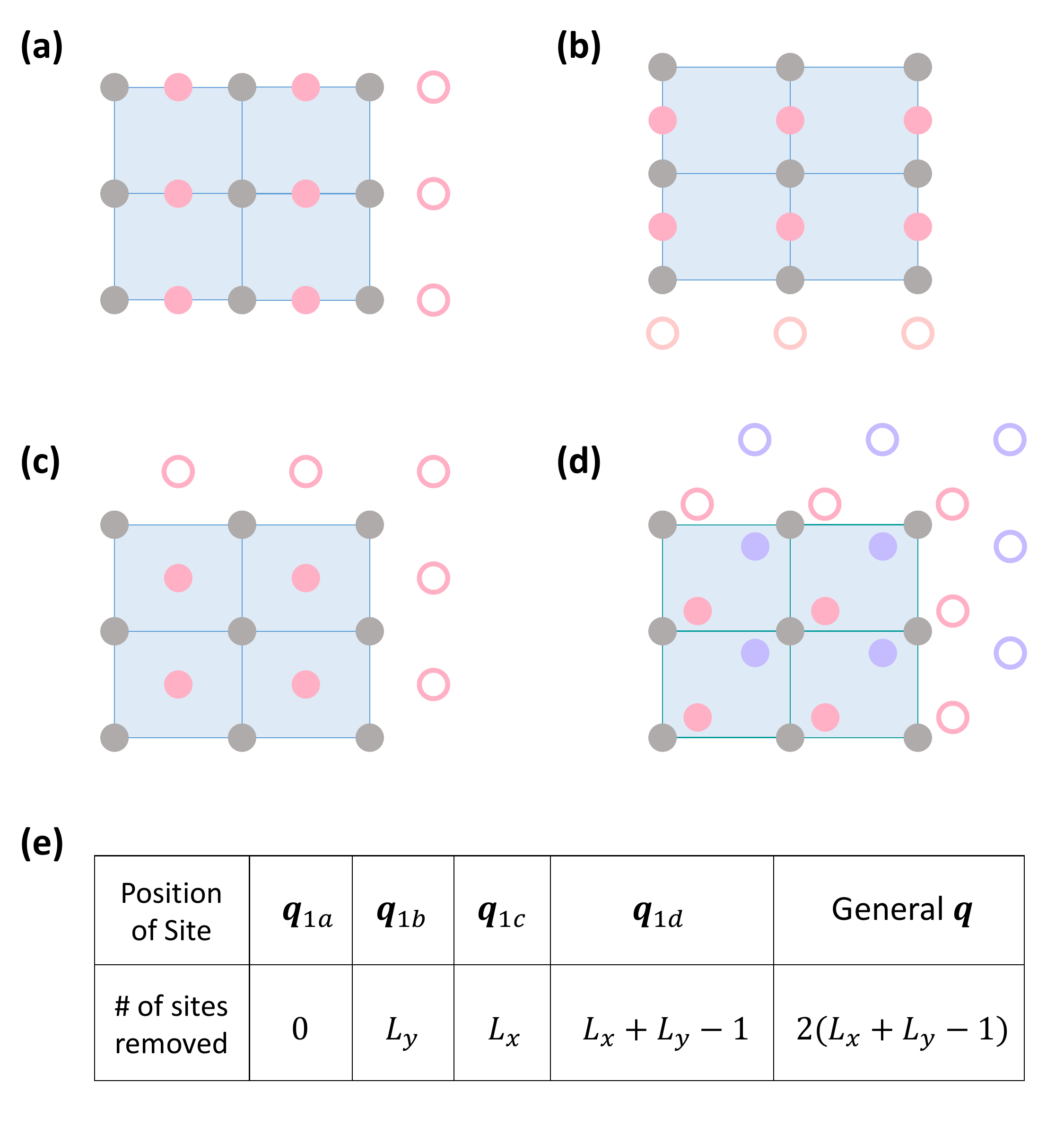}
	\caption{The solid dots form inversion-symmetric open lattice geometries for atomic site configurations in (a) - (d). Boundary atomic sites that need to be eliminated to preserve inversion symmetry are shown in colored circles. In (e), we list the number of atomic sites that need to be removed for an open geometry with $L_x \times L_y$ unit cells if we assign an atomic site to a specific Wyckoff position.}
	\label{Fig: Atomic Sites}
\end{figure}

Now let us elaborate on another key ingredient for understanding the higher-order topology in Kitaev superconductors, which is the role of atomic sites in defining an inversion-symmetric open boundary. With the Kitaev building blocks as a complete set of basis, we will show that the choice of symmetric open boundary is closely related to how we ``stack" the Kitaev building blocks.

In principle, there are two distinct stacking strategies with the Kitaev building blocks to generate all possible Kitaev superconductors:

\begin{itemize}
	\item {\it Face-to-face stacking}: The atomic sites of all stacked building block coincide with each other in real space at ${\bf q}_{1a}$.
	\item {\it Displaced stacking}: The atomic sites of some stacked building blocks are different from ${\bf q}_{1a}$.  
\end{itemize}
As a reference point, we always assume the existence of at least one building block, whose atomic site coincides with the origin ${\bf q}_{1a}$. For the displaced stacking, when we place the atomic site of an Kitaev building block $\kappa_i$ at some general/maximal Wyckoff position ${\bf q}$, we also put its corresponding Wannier orbital at ${\bf q} + {\bf q}_{1j}$ in order to make the building block well-defined.

After the face-to-face stacking, the composite system has no sublattice degree of freedom. Since the inversion center is always set at the origin ${\bf q}_{1a}$ in our convention, any rectangular open geometry with an odd-integer number of unit cells along $x$ and $y$ directions would preserve the inversion symmetry.

For the displaced stacking, however, the collection of atomic sites within one single unit cell are not invariant under its inversion center as a whole. Therefore, in an open geometry with an integer number of unit cells, we will inevitably eliminate a fraction of atomic sites on the boundary to preserve the global inversion. For illutration, we plot in Fig. \ref{Fig: Atomic Sites} four inequivalent displaced stacking configurations. For an open geometry with $L_x\times L_y$ unit cells ($L_{x,y} \in$ odd), if we place one atomic site at ${\bf q}_{1a}$ and another at ${\bf q}_{1b}$ (${\bf q}_{1c}$), the number of atomic sites being disregarded (shown in red circle) is exactly $L_y$ ($L_x$), as shown in Fig. \ref{Fig: Atomic Sites} (a) and (b). If we place one atomic site at ${\bf q}_{1a}$ and another at ${\bf q}_{1d}$, the number of sites being ignored is actually $L_x+L_y-1$, with this extra $-1$ coming from the corner atomic site shared by the $x$ and $y$ edges, as shown in Fig. \ref{Fig: Atomic Sites} (c). On the other hand, if an atomic site (red dot) is placed at a non-maximal Wyckoff position ${\bf q}$, there must be another atomic site (purple dot) at the inversion-related position ${\cal I}{\bf q}$ to preserve the inversion symmetry, as shown in Fig. \ref{Fig: Atomic Sites} (d). In this case, the number of removed atomic sites on the boundary is $2(L_x+L_y-1)$. To summarize, we have listed the results in Fig. \ref{Fig: Atomic Sites} (e). \\

The necessity of removing residue atomic sites for the displaced-stacking models originates from their inversion-breaking unit cells. While the choice of unit cell is practically flexible, we notice that when no atomic site sits on a maximal Wyckoff position ${\bf q}_{1i} \neq {\bf q}_{1a}$ [as shown in Fig. \ref{Fig: Atomic Sites} (d)], it is always possible to find an inversion-invariant choice of unit cell. However, when at least one atomic site sits on a maximal Wyckoff position ${\bf q}_{1i} \neq {\bf q}_{1a}$ [as shown in Fig. \ref{Fig: Atomic Sites} (a) - (c)], all choices of the unit cell necessarily break inversion symmetry. The latter case exactly corresponds to our displaced stacking strategy.

Crucially, an inversion-breaking unit cell guarantees that the following two conditions in momentum space {\it cannot be fulfilled simultaneously}: 
\begin{enumerate}
	\item[(i)] the matrix representation of inversion symmetry is independent of crystal momentum ${\bf k}$;
	\item[(ii)] the Hamiltonian is $2\pi$-periodic in ${\bf k}$.
\end{enumerate}	
In principle, whether a Hamiltonian is $2\pi$-periodic in ${\bf k}$ depends on the explicit definition of Fourier transformation that switch between a real-space basis to a momentum-space one. Therefore, if we require a $2\pi$-periodic momentum-space Hamiltonian, then the corresponding inversion representation should generally be ${\bf k}$-dependent when its unit cell does not respect inversion symmetry. In Appendix \ref{Appendix A}, we provide several examples to explicitly show this deep connection between the inversion-breaking unit cell and the ${\bf k}$-dependence of inversion representation.

\subsection{The Majorana Counting Rule: A Derivation} \label{Sec: Deriving Counting Rule}

Now we are ready to derive the bulk Majorana counting rule for Kitaev superconductors. 

As shown in Fig. \ref{Fig: Atomic Polarization}, we first notice that the number of dangling Majorana modes for each Kitaev building blocks is exactly half the number of dangling Wannier orbitals in its Wannier representation. This is simply because the Wannier orbitals in Fig. \ref{Fig: Atomic Polarization} are only shown for the occupied states, while their particle-hole partners for the unoccupied states exactly sit at the same location. Similar to a Kitaev chain, such particle-hole pair of Wannier orbitals on the boundary essentially originate from a non-local superposition of one $\alpha$-type Majorana fermion and one $\beta$-type Majorana fermion. This completes the mapping between the Majorana representation and the Wannier representation for a general Kitaev limit with open boundary conditions.  

Recall that in Fig. \ref{Fig: Atomic Polarization}, we also list $(N_x,N_y)$ for each building block with $L_x$ by $L_y$ unit cells. So one might naively expect that, ${\cal N_I}(\kappa_i)$, the number of boundary Majorana modes modulo ${\cal I}$ for a building block $\kappa_i$, is simply $N_x+N_y$. However, we will then have missed a crucial fact that some boundary atomic sites must be removed along with their Wannier orbitals to preserve the inversion symmetry, as we have discussed in Sec. \ref{Sec: Atomic Sites}.    

To implement a correct counting of boundary Majorana modes, we consider a general Kitaev limit with $n_i^{(A)}$ atomic sites and $n_i^{(W)}$ Wannier orbitals at maxiaml Wyckoff position ${\bf q}_{1i}$. On an open geometry with complete $L_x\times L_y$ unit cells, we have ${\cal N_I}^{(0)} = L_y n_b^{(W)} + L_x n_c^{(W)} + (L_x+L_y-1) n_d^{(W)}$ Majorana modes dangling on the boundary. On the other hand, following Fig. \ref{Fig: Atomic Sites} (e), we also need to remove ${\cal N_I}^{(1)} = L_y n_b^{(A)} + L_x n_c^{(A)} + (L_x+L_y-1) n_d^{(A)}$ numbers of atomic sites along with their Wannier orbitals (or equivalently boundary Majorana modes). All these countings together lead to 
\begin{eqnarray}
	{\cal N_I}  = L_x (\Delta_c + \Delta_d) + L_y (\Delta_b + \Delta_d) - \Delta_d, 
	\label{Eq: counting number N_I}
\end{eqnarray}
where we have defined the Majorana counting number $\Delta_i = n_b^{(W)}-n_b^{(A)}$. Since we only care about the oddness of ${\cal N_I}$, we will not count any contributions from both atomic sites and Wannier orbitals at a non-maximal Wyckoff position, which only contribute evenly to ${\cal N_I}$. 

On the other hand, we hope to rule out the possibility of weak TSCs by imposing a polarization constraint ${\cal P} = 0$. It is then straightforward to show that a vanishing ${\cal P}$ is equivalent to
\begin{eqnarray}
	\Delta_b + \Delta_d \equiv \Delta_c + \Delta_d \equiv 0\ \ (\text{mod }2).
\end{eqnarray}
Together with Eq. \ref{Eq: counting number N_I}, we find that the condition for hosting higher-order topology with an odd ${\cal N_{B_I}}$ is only possible when $\Delta_{b,c,d}$ are all odd. This is exactly the Majorana counting rule that we have stated in Sec. \ref{Sec: Majorana counting rule}. Remarkably, this simple counting rule only relies on the position information of atomic sites and Wannier orbitals.

In the following section, we will construct explicit examples to demonstrate both the stacking-based construction of Kitaev superconductors with higher-order topology and how they can be correctly diagnosed via the Majorana counting rule. 

\section{Building-block Construction of Higher-order TSCs} \label{Sec: Examples}

In this section, we propose three stacking recipes to constructed higher-order TSC phases based on the Kitaev building blocks, all of which are exemplified by minimal model constructions. The recipes are directly motivated by the  Majorana counting rule.  

\subsection{Face-to-face Stacking} \label{Sec: Face-to-face Stacking}

Let us first focus on the face-to-face stacking strategy. Starting from the Majorana counting rule, we first derive a simple stacking recipe that necessarily leads to higher-order topology. We will then construct a minimal face-to-face stacking model with higher-order topology that demonstrate both the stacking reciple and the counting rule.  

\subsubsection{Recipe of Face-to-face Stacking}

For face-to-face stacking of the Kitaev building blocks, by definition, all the atomic sites sit at the unit-cell origin ${\bf q}_{1a}$. Therefore, if we stack $n_i$ copies of building block $\kappa_i$ with $i=a,b,c,d$, the counting numbers are given by
\begin{eqnarray}
	\Delta_a &=& - (n_b + n_c + n_d),\  \Delta_b = n_b,\nonumber \\
	\Delta_c &=& n_c,\ \Delta_d = n_d.
\end{eqnarray}
According to the Majorana counting rule, we immediately arrive at the following stacking recipe for higher-order TSC:

\begin{itemize}
	\item {\bf Stacking Recipe $\#1$}: Face-to-face stacking $n_i$ copies of building block $\kappa_{i}$ ($i=a,b,c,d$) will lead to higher-order topology {\it if and only if} $n_{b,c,d}\in$ odd.
\end{itemize}

According to this recipe, the minimal face-to-face stacking model that hosts higher-order topology is to choose $n_b=n_c=n_d=1$. In the following, we will construct such minimal tight-binding model and explicitly demostrate the existence of corner-localized Majorana zero modes.  

\subsubsection{A Minimal Face-to-face Stacking Model with Higher-order Topology}

\begin{figure*}[t]
	\centering
	\includegraphics[width=0.98\textwidth]{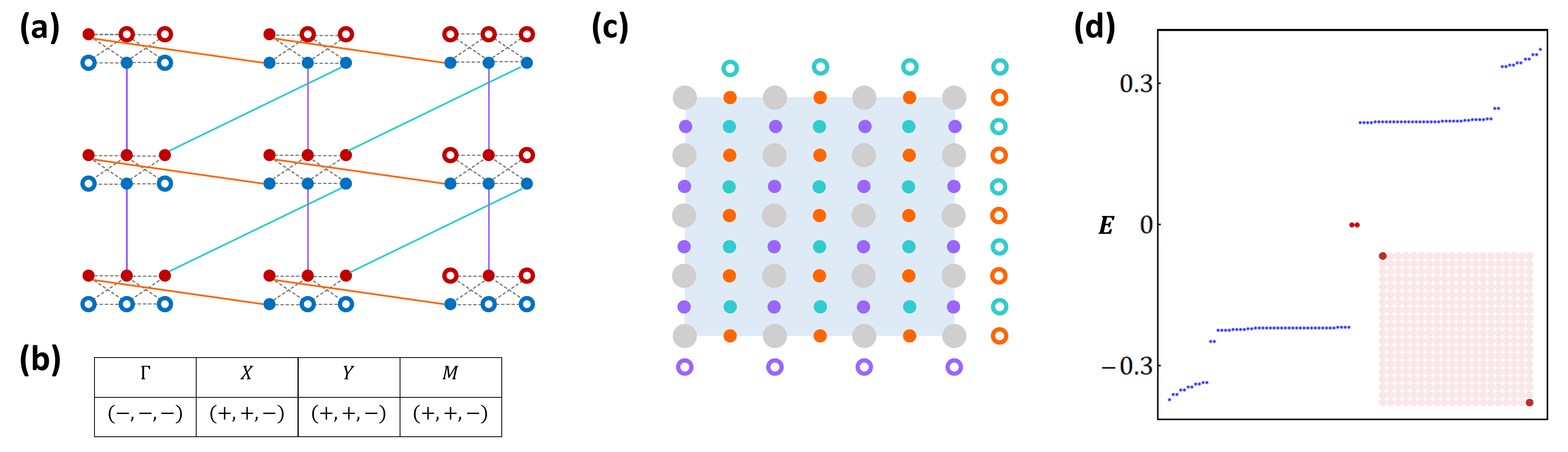}
	\caption{Our minimal model for a nontrivial face-to-face stacking in the Majorana representation is shown (a). The red and blue dots denote $\alpha$ and $\beta$ Majorana fermions, while the colored circles in (a) denote unpaired boundary Majorana fermions for $h_{sb}$. In particular, the colored solid (black dashed) lines denote the strong Majorana bonds in $h_{sb}$ (weak Majorana bonds in $h_{wb}$). The parity data for each high-symmetry momentum is shown in (b). We plot in (c) the distribution of BdG Wannier orbitals, with the large grey dots for the atomic sites and the small colored dots (or circles) for the Wannier orbitals. (d) shows the energy spectrum of $H_1({\bf k})$ in an open geometry. There exists a pair of Majorana zero-energy modes (highlighted in red) that are localized on the top-left and bottom-right corners, as shown in the inset of (d).}
	\label{Fig: Face to face}
\end{figure*}

We consider three electron orbitals labeled by an orbital index $l=b,c,d$ at Wyckoff position ${\bf q}_{1a}$, as well as the corresponding hole partners. The electron annihilation and creation operators can be written as a linear superposition of Majorana operators $\alpha_{{\bf R},l}$ and $\beta_{{\bf R},l}$ as
\begin{equation}
c_{{\bf R},l} = \frac{\alpha_{{\bf R},l} + i \beta_{{\bf R},l} }{\sqrt{2}},\ 
c_{{\bf R},l}^{\dagger} = \frac{\alpha_{{\bf R},l} - i \beta_{{\bf R},l} }{\sqrt{2}}.
\end{equation} 
A schematic plot of the lattice configuration in the Majorana representation is shown in Fig. \ref{Fig: Face to face} (a), where the red dots and the blue dots denote $\alpha$-type and $\beta$-type Majorana degrees of freedom, respectively. 

To realize the face-to-face stacking, we introduce the following inter-unit-cell Majorana bonds shown in Fig. \ref{Fig: Face to face} (a): 
\begin{enumerate}
	\item[(i)] the red bond $t_1$ connects $\alpha_{{\bf R},b}$ and $\beta_{{\bf R}+{\bf a}_x,b}$, which realizes a copy of $\kappa_b$ with ${\cal P}=(\frac{1}{2},0)$;
	\item[(ii)] the purple bond $t_2$ connects $\alpha_{{\bf R},c}$ and $\beta_{{\bf R}+{\bf a}_y,c}$, which realizes a copy of $\kappa_c$ with ${\cal P}=(0,\frac{1}{2})$;
	\item[(iii)] the green bond $t_3$ connects $\alpha_{{\bf R},d}$ and $\beta_{{\bf R}+{\bf a}_x+{\bf a}_y,d}$, which realizes a copy of $\kappa_d$ with ${\cal P}=(\frac{1}{2},\frac{1}{2})$.
\end{enumerate}
As a demonstration, the configuration of BdG Wannier orbitals is schematically plotted in Fig. \ref{Fig: Face to face} (c), where the color of each Wannier orbital matches with that of its corresponding Majorana bond. 

While the colored Majorana bonds necessarily gap out the bulk states, there are still some unwanted dangling Majorana modes (or equivalently dangling Wannier orbitals) on the edge, as shown by the open circles in Fig. \ref{Fig: Face to face} (a) and (c). To further remove these in-gap edge degree of freedom, we introduce additional {\it weak} Majorana bonds among different species of Majorana modes within one unit cell, as shown by the black dashed lines. Therefore, in terms of Majorana operators, the Hamiltonian consists of $h_{sb}$ describing the colored strong bonds and $h_{wb}$ for the dashed weak bonds. In the Majorana representation, we have
\begin{widetext}
	\begin{eqnarray}
	h_{sb} &=& i\sum_{\bf R} [t_1 \alpha_{{\bf R},b}\beta_{{\bf R}+{\bf a}_x,b} + t_2 \alpha_{{\bf R},c}\beta_{{\bf R}+{\bf a}_y,c} + t_3 \alpha_{{\bf R},d}\beta_{{\bf R}+{\bf a}_x + {\bf a}_y,d}], \nonumber \\
	h_{wb} &=& \sum_{\bf R} [i u_{12} (\alpha_{{\bf R},b} \alpha_{{\bf R},c} + \beta_{{\bf R},b} \beta_{{\bf R},c}) + i u_{23} (\alpha_{{\bf R},c} \alpha_{{\bf R},d} + \beta_{{\bf R},c} \beta_{{\bf R},d}) + iv_{12} (\alpha_{{\bf R},b} \beta_{{\bf R},c} + \alpha_{{\bf R},c} \beta_{{\bf R},b})  \nonumber \\
	&& + i v_{23} (\alpha_{{\bf R},c} \beta_{{\bf R},d} +  \alpha_{{\bf R},d} \beta_{{\bf R},c})].
	\end{eqnarray}   
\end{widetext}
In the momentum space, we consider the following BdG basis
\begin{equation}
\Psi({\bf k}) = (c_{{\bf k},b}, c_{{\bf k},c}, c_{{\bf k},d},c_{-{\bf k},b}^{\dagger}, c_{-{\bf k},c}^{\dagger}, c_{-{\bf k},d}^{\dagger})^T,
\end{equation}
and the above Hamiltonians become
\begin{widetext}
	\begin{eqnarray}
	H_1 ({\bf k}) = \begin{pmatrix}
	t_1 \cos k_x & i u_{12} + v_{12} & 0 & it_1 \sin k_x & 0 & 0 \\
	-i u_{12} + v_{12} & t_2 \cos k_y & i u_{23} + v_{23} & 0 & i t_2 \sin k_y & 0 \\
	0 & -i u_{23} + v_{23} & t_3 \cos (k_x + k_y) & 0 & 0 & i t_3\sin (k_x + k_y) \\
	-i t_1 \sin k_x & 0 & 0 & -t_1 \cos k_x & i u_{12} - v_{12} & 0 \\
	0 & -i t_2\sin k_y & 0 & -i u_{12} - v_{12} & -t_2\cos k_y & i u_{23} + v_{23} \\
	0 & 0 & -it_3\sin (k_x+k_y) & 0 & -i u_{23} + v_{23} & -t_3\cos (k_x+k_y) \\
	\end{pmatrix}
	\end{eqnarray}
\end{widetext}
The inversion operator for the bulk Hamiltonian $H_1({\bf k})$ is given by ${\cal I} = \tau_z \otimes \mathbb{1}_3$. Here $\tau_z$ is a Pauli matrix for the particle-hole index and $\mathbb{1}_3$ is a $3\times 3$ identity matrix characterizing the orbital index $l$. For our choice of parameters, it is easy to check that the parity data for the occupied bands at each high-symmetry momenta is given by Fig. \ref{Fig: Face to face} (b). Notably, such parity data directly implies a ``double band inversion" at $\Gamma$ point \cite{hsu2019inversion} and agrees with a prediction of higher-order TSC from symmetry indicators \cite{ono2019symmetry,skurativska2020atomic,ono2019refined,geier2019symmetry}. 

To confirm $H_1({\bf k})$ as a higher-order TSC, we calculate its energy spectrum on a square lattice with open boundary conditions in both $x$ and $y$ directions. As shown in Fig. \ref{Fig: Face to face} (d) and its inset, we indeed find a pair of corner-localized Majorana zero modes, which unambiguously demonstrate the inversion-protected higher-order topology in our system. We note that the corner Majorana modes in $H_f$ has vanishing localzation length, which is similar to the Kitaev limit of a 1d Majorana chain. This numerical result agrees with our Majorana counting rule and the stacking recipe \#1.

\subsection{Displaced Stacking with the Same Building Blocks} \label{Sec: displaced stacking H2}

To go beyond the face-to-face stacking, we now turn to a more general situation where the building blocks are stacked in a displaced way. Specifically, the atomic sites of the building blocks do NOT have to coincide at ${\bf q}_{1a}$. This stacking strategy inspires us to define a displacement vector ${\bf d}=(d_x,d_y)$ to characterize the displacement between the atomic site and the origin ${\bf q}_{1a}$. Since only the atomic sites and Wannier orbitals sitting at maximal Wyckoff positions can contribute to the higher-order topology, the components of ${\bf d}$ must be half-integer-valued with respect to the lattice constant with  $d_{x,y}\in\{0,\frac{1}{2}\}$. From now on, we will denote that a building block is sitting at a Wyckoff position ${\bf q}$ if its atomic site coincides with ${\bf q}$.\\

Let us first start with the displaced stacking construction with {\it only} one kind of building block, i.e. $n_i$ copies of building block $\kappa_i$ ($i=b,c,d$). To avoid Majorana edge band, we first require $n_i\in$ even to trivialize bulk polarization ${\cal P}$. In addition, we will ignore the case where an even number of building blocks of the same kind share the same atomic site, because such stacked system is always topologically trivial. As a result, it is sufficient to consider a simple double-stacking model with $n_i=2$, where one building block $\kappa_i$ sits at ${\bf q}_{1j_1}$ and another building block $\kappa_i$ sits at ${\bf q}_{1j_2}$ for $j_1\neq j_2\in\{a,b,c,d\}$. Such double-stacking model is uniquely characterized by a {\it relative} displacement vector $\delta {\bf d}^{(i)} = {\bf q}_{1j_2} - {\bf q}_{1j_1}$, where the superscript $i$ labels the type of building block.

For the double-stacking models, we have identified a necessary and sufficient condition for the presence of higher-order topology, which can be easily tested.
\begin{itemize}
	\item {\bf Stacking Recipe \#2}: A double-stacking system with building block $\kappa_i$ is higher-order topological {\it if and only if} the relative displacement vector $\delta {\bf d}^{(i)} \not\in \{{\bf q}_{1a},{\bf q}_{1i}\}$ for $i\in \{ b,c,d\}$.
\end{itemize}

For example, a topologically trivial example can be constructed by placing one $\kappa_b$ at ${\bf q}_{1a}$ and another $\kappa_b$ at ${\bf q}_{1b}$ such that $\delta {\bf d}^{(b)} = {\bf q}_{1b}$. This can be understood from the Majorana counting rule, since we now have an equal number of atomic sites and Wannier orbitals for all maximal Wyckoff positions and consequently all Majorana counting numbers are zero. 

However, if we move the $\kappa_b$ at ${\bf q}_{1b}$ to ${\bf q}_{1c}$, the system is then expected to be higher-order topological following the recipe, since $\delta {\bf d}^{(b)} = {\bf q}_{1c}$. From the counting perspective, we have one atomic site (Wannier orbital) at ${\bf q}_{1a}$ and ${\bf q}_{1c}$ (${\bf q}_{1b}$ and ${\bf q}_{1d}$), respectively, which directly indicates nontrivial counting numbers
\begin{equation}
	\Delta_a = -\Delta_b = \Delta_c = -\Delta_d = -1 
\end{equation}
In the remaining part of this subsection, we will check this prediction by constructing a corresponding minimal double-stacking model and explicitly demonstrating the existence of corner-localized Majorana zero modes.   

\begin{figure*}[t]
	\centering
	\includegraphics[width=0.99\textwidth]{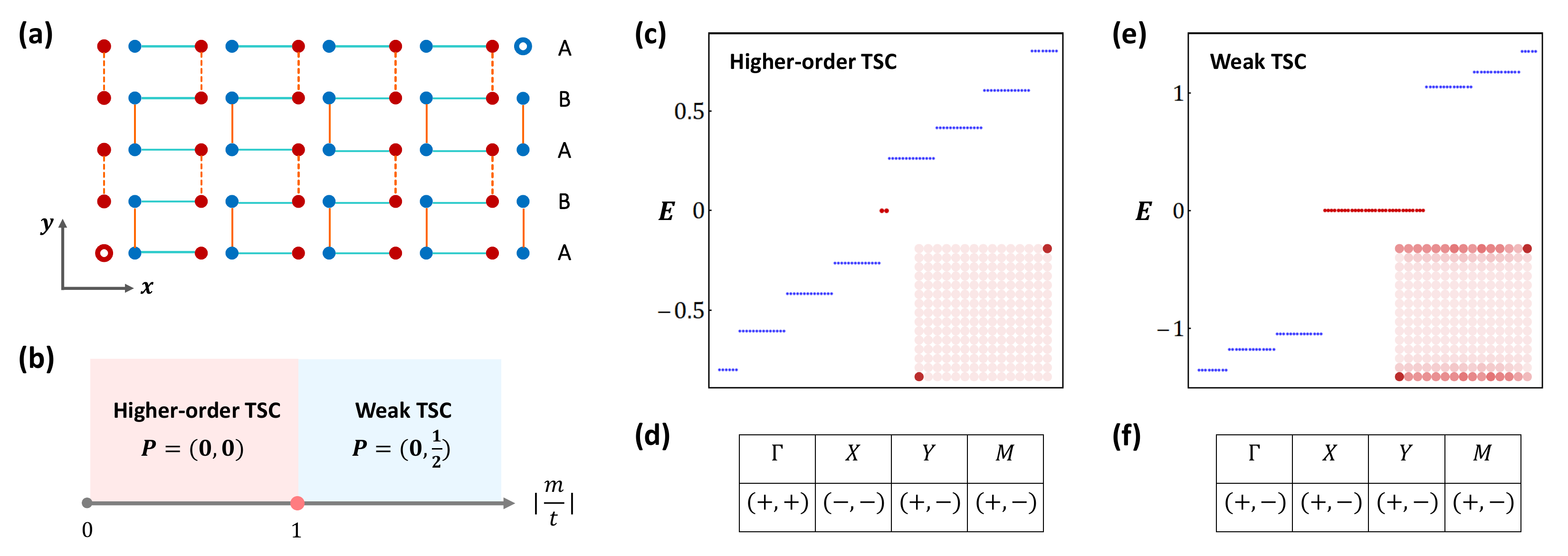}
	\caption{The Majorana representation of the minimal double-stacking model $H_2({\bf k})$ is shown in (a). The topological phase diagram of $H_2$ as a function of $|m/t|$ is shown in (b). We plot the energy spectrum for the higher-order TSC phase on an open geometry in (c) and list its parity data in (d). The energy spectrum and the parity data for the weak TSC phase with ${\cal P}=(0,\frac{1}{2})$ are shown in (e) and (f), respectively. }
	\label{Fig: Dimer1}
\end{figure*}

\subsubsection{A Minimal Double-Stacking Model with Higher-order Topology}

The minimal double-stacking model with a relative displacement vector $\delta {\bf d}^{(b)} = {\bf q}_{1c}$ is schematically shown in Fig. \ref{Fig: Dimer1} (a), where red and blue dots denote $\alpha$-type and $\beta$-type Majorana operators.  The lattice constants ${\bf a}_x$ and ${\bf a}_y$ are also shown in Fig. \ref{Fig: Dimer1} (a). For our purpose, we include two pairs of Majorana modes within one unit cell, with one pair at ${\bf r}_A={\bf q}_{1a}=(0,0)$ with a sublattice index $A$ and another pair at ${\bf r}_B={\bf q}_{1c}=(0,\frac{1}{2})$ with an index $B$. The Majorana operators are related to the electron creation and annihilation operators as 
\begin{equation}
c_{{\bf R},l} = \frac{\alpha_{{\bf R},l} + i \beta_{{\bf R},l} }{\sqrt{2}},\ 
c_{{\bf R},l}^{\dagger} = \frac{\alpha_{{\bf R},l} - i \beta_{{\bf R},l} }{\sqrt{2}},
\end{equation}
where $l=A,B$ is the sublattice index and ${\bf R}$ is the lattice vector that defines the unit cell.

Three types of Majorana bonds are considered in this minimal model. As shown in Fig. \ref{Fig: Dimer1} (a),
\begin{enumerate}
	\item[(i)] the green bond $t$ connects $\beta_{{\bf R},l}$ and $\alpha_{{\bf R} + {\bf a}_x, l}$, which realizes a copy of $\kappa_b$ for individual sublattice if they are the strongest bonds;
	\item[(ii)] the dashed orange bond $m_{\alpha}$ connects $\alpha_{{\bf R},B}$ and $\alpha_{{\bf R}+{\bf a}_y,A}$;
	\item[(iii)] the solid orange bond $m_{\beta}$ connects $\beta_{{\bf R},A}$ and $\beta_{{\bf R},B}$.
\end{enumerate}
Then the model Hamiltonian in the Majorana representation is given by
\begin{eqnarray}
	H_2({\bf k}) &=& it \beta_{{\bf R},l} \alpha_{{\bf R}+{\bf a}_x,l} + im_{\alpha} \alpha_{{\bf R},B} \alpha_{{\bf R}+{\bf a}_y,A} \nonumber \\
	&& + im_{\beta} \beta_{{\bf R}+{\bf a}_y,B} \beta_{{\bf R},A}.
\end{eqnarray}
The inversion operation will switch between $\alpha$ and $\beta$ as 
\begin{equation}
	{\cal I} \begin{pmatrix}
	\alpha_{{\bf R},l} \\
	\beta_{{\bf R},l}
	\end{pmatrix} = 
	\begin{pmatrix}
	0 & -1 \\
	1 & 0  \\
	\end{pmatrix}
	\begin{pmatrix}
	\alpha_{{\bf R}',l} \\
	\beta_{{\bf R}',l}
	\end{pmatrix},
\end{equation}
where ${\bf R}$ and ${\bf R}'$ are related by inversion symmetry. It is easy to check that inversion symmetry will enforce $m_{\alpha} = m_{\beta}=m$. 

Back to the fermion basis, we consider the Fourier transformation
\begin{equation}
	c_{{\bf R},l} = \sum_{\bf k} e^{i{\bf k}\cdot ({\bf R} + {\bf r}_l)} c_{{\bf k},l},
	\label{Eq: Fourier transform}
\end{equation}
and the momentum-space Hamiltonian is 
\begin{eqnarray}
	&& H_2 ({\bf k}) \nonumber \\
	 &=& \begin{pmatrix}
	 -t \cos k_x & -im \cos \frac{k_y}{2} & it \sin k_x & m\sin \frac{k_y}{2} \\
	 im \cos \frac{k_y}{2} & -t \cos k_x & m\sin \frac{k_y}{2} & it \sin k_x \\
	 -it \sin k_x & m\sin \frac{k_y}{2} & t \cos k_x & -im \cos \frac{k_y}{2} \\
	 m\sin \frac{k_y}{2} & -it \sin k_x & im \cos \frac{k_y}{2} & t \cos k_x \\
	\end{pmatrix}. \nonumber \\
	&&
	\label{Eq: H2(k)}
\end{eqnarray}
in the Nambu basis
\begin{equation}
	\Psi({\bf k}) = (c_{{\bf k},A},c_{{\bf k},B},c_{-{\bf k},A}^{\dagger},c_{-{\bf k},B}^{\dagger})^T.
\end{equation} 
We note that $H_x^{(d)}({\bf k})$ can be diagonalized analytically with eigen-energy as
\begin{equation}
	E = \pm \sqrt{m^2+t^2 \pm 2mt\cos \frac{k_y}{2}}.
\end{equation}
Therefore, there exists a topological phase transition (bulk gap closing) at $|t|=|m|$ that separate two distinct phases. 

\subsubsection{$H_2({\bf k})$ with $|t|>|m|$: A Higher-order TSC} \label{Subsec: HOTSC in H2}
When $|t|>|m|$, we consider the Kitaev limit by setting $|m|\rightarrow 0$. Then the BdG Wannier orbitals of the double-stacking model are exactly localized at the center of the green Majorana bonds. Therefore, the system is equivalent to a stacking of one $\kappa_b$ at ${\bf q}_{1a}$ and another $\kappa_{b}$ at ${\bf q}_{1c}$, which is featured by a relative displacement vector $\delta {\bf d}^{(b)} = {\bf q}_{1c}$. According to our stacking recipe \#2, this phase is diagnosed to be a higher-order TSC.     

To confirm its higher-order nature, we numerically calculate the energy spectrum of this double-stacking model with $t=1.2$ and $m=1$ on an open geometry. To preserve the inversion symmetry, it is crucial to choose only the atomic sites with a sublattice index $l=A$ as the edge termination for both the upper and the lower $y$-edges, similar to the lattice geometry in Fig. \ref{Fig: Dimer1} (a). Just as we expect, the energy spectrum in Fig. \ref{Fig: Dimer1} (c) hosts a pair of Majorana modes exactly at zero energy. In the inset of Fig. \ref{Fig: Dimer1} (c), we further plot the wavefunction distribution in the real space for both Majorana modes. Sitting in the opposite corners of the lattice, these Majorana modes are indeed corner-localized and unambiguously signals the inversion-protected higher-order topology.  

In Fig. \ref{Fig: Dimer1} (d), we further list the parity data of this higher-order TSC phase at each high-symmetry momentum. To correctly calculate the parity data, we need to modify the Fourier transformation in Eq. \ref{Eq: Fourier transform} to $c_{{\bf R},l} = \sum_{\bf k} e^{i{\bf k}\cdot {\bf R}} c_{{\bf k},l}$. Then the inversion operator is found to be ${\bf k}$-dependent with
\begin{equation}
	{\cal I} = \begin{pmatrix}
	1 & & & \\
	& e^{i k_y} & & \\
	& & -1 & \\
	& & & -e^{ik_y} \\
	\end{pmatrix}.
\end{equation}
As we have discussed in Sec. \ref{Sec: Atomic Sites}, the ${\bf k}$-dependence of ${\cal I}$ originates from the displaced stacking.

To understand the parity data intuitively, we first note that the system is essentially an array of $x$-directed Kitaev chain in the Kitaev limit $m\rightarrow 0$. Thus, the system has only one occupied band and the parity data is simply 
\begin{equation}
[(+)_{\Gamma}, (-)_X, (+)_Y, (-)_M].
\end{equation} 
Here $(+)_{\Gamma}$ denotes one positive parity for the occupied band at $\Gamma$ point. By turning on $m$, the system becomes dimerized along $y$-direction and leads to Brillouin zone (BZ) folding in the momentum space. As a result, the parity data at $Y$ and $M$ will be folded to $\Gamma$ and $M$, respectively. We thus have $(+,+)_{\Gamma}$ and $(-,-)_{X}$. Meanwhile, $\tilde{Y}=(0,\frac{\pi}{2})$ transforms into $\tilde{Y}'=(0,-\frac{\pi}{2})$ under inversion, which is thus not an inversion-invariant momentum {\it before} the BZ folding. When the BZ folding happens, the BdG states at $\tilde{Y}'$ and $\tilde{Y}$ are folded together as the new $Y$ point and the inversion operation switches between them as a $\sigma_x$ operation. Therefore, the inversion eigenstates at $Y$ after the BZ folding are necessarily the bonding and anti-bonding combinations of the states that are originally from $\tilde{Y}$ and $\tilde{Y}'$. As a result, the occupied bands at $Y$ must host an equal number of ``$+$" and ``$-$" parity eigenvalues after the BZ folding along $k_y$ direction. A similar argument can be applied to the parity data at $M$. Therefore, we arrive at the following parity data for the higher-order TSC phase 
\begin{equation}
[(+,+)_{\Gamma}, (-,-)_X, (+,-)_Y, (+,-)_M],
\end{equation} 
as shown in Fig. \ref{Fig: Dimer1} (c).

\subsubsection{$H_2({\bf k})$ with $|t|<|m|$: A Weak TSC}

When $|t|<|m|$, we take the Kitaev limit by setting $|t| \rightarrow 0$ instead. Then the Wannier orbitals will sit at the center of the orange bonds $m$. To be specific, the system has one Wannier orbital at a non-maximal Wyckoff position ${\bf q}=(0,\frac{1}{4})$ and another at ${\bf q}=(0,\frac{3}{4})$. On the other hand, the two atomic sites sit at ${\bf q}_{1a}$ and ${\bf q}_{1c}$, respectively. According to the Majorana counting rule, we have
\begin{equation}
	\Delta_a = \Delta_c = -1,\ \Delta_b = \Delta_d = 0,
\end{equation}
and this phase is thus NOT higher-order topological. 

In fact, the two Wannier orbitals inside one unit cell are inversion-symmetric around ${\bf q}_{1a}$. As a result, we can simultaneously move both Wannier orbitals to ${\bf q}_{1a}$ in a symmetric and adiabatic way. Then the system is equivalent to a stacking of a $\kappa_a$ at ${\bf q}_{1a}$ and a $\kappa_c$ at ${\bf q}_{1c}$, which does not satisfy higher-order condition in our recipe \#2.  
 
By counting the relative distance between atomic sites and the Wannier orbitals, we identify this phase as a weak TSC with a nontrivial polarization ${\cal P} = (0,\frac{1}{2})$. To verify this, we calculate the energy spectrum of the weak TSC phase with open boundary conditions in both directions. As shown in Fig. \ref{Fig: Dimer1} (d) and its inset, this weak TSC phase indeed hosts a single Majorana flat band localized on its $y$-edge, as predicted by its polarization. We also show its parity data in Fig. \ref{Fig: Dimer1} (e).

\subsection{Displaced Stacking with Different Building Blocks}\label{Sec: Beyond indicator}

We can also stack different Kitaev building blocks in a displaced way to achieve higher-order topology. While it is difficult to prove a necessary and sufficient higher-order topological recipe for displaced stackings with inequivalent building blocks, we do emphasize that our Majorana counting rule still holds for diagnosing the higher-order topology. Hence, in the following, we will only provide a minimal example to demonstrate this stacking strategy and leave the proposal of a general recipe for future works.

\begin{figure*}[t]
	\centering
	\includegraphics[width=0.93\textwidth]{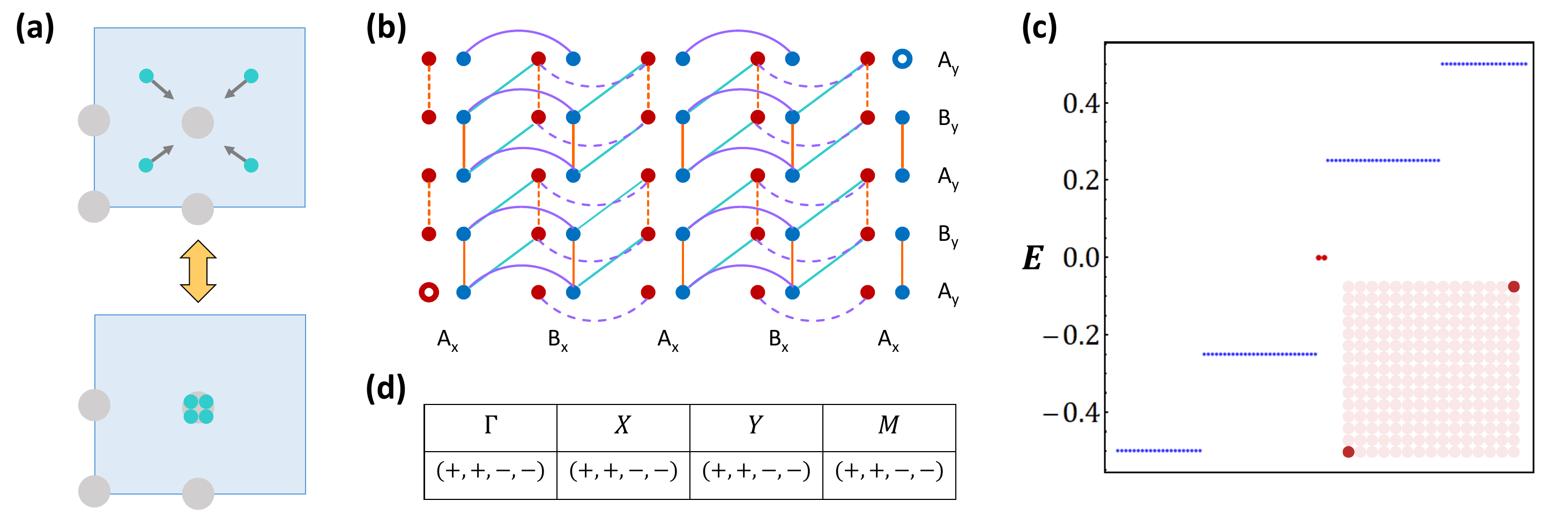}
	\caption{Higher-order topology of displaced stacking model $H_3$. In (a), we show the adiabatic process by symmetrically moving Wannier orbitals at different general Wyckoff positions to a maximal Wyckoff position. The Majorana representation of $H_3$ is shown in (b). The higher-order topology of $H_3$ is confirmed in (c) and its inset by plotting the energy spectrum on an open geometry. The parity data for $H_3$ is shown in (d).}
	\label{Fig: dimer2}
\end{figure*}

\subsubsection{Model Hamiltonian and Higher-order Topology}

We now consider a displaced stacking with all four inequivalent building blocks: (i) one $\kappa_a$ at ${\bf q}_{1d}$, (ii) one $\kappa_b$ at ${\bf q}_{1c}$, (iii) one $\kappa_c$ at ${\bf q}_{1b}$, and (iv) one $\kappa_d$ at ${\bf q}_{1a}$. In this case, we have four atomic sites distributed at all four maximal Wyckoff positions and four Wannier orbitals coincide at ${\bf q}_{1d}$. Following the counting rule, we have
\begin{equation}
	\Delta_a = \Delta_b = \Delta_c = -1,\ \Delta_d = 3,
\end{equation}
and the system is predicted to host higher-order topology.

To confirm this prediction, we consider a topologically equivalent model $H_3$ by symmetrically moving the Wannier orbitals at ${\bf q}_{1d}$ to four non-maximal positions: $(\frac{1}{4},\frac{1}{4}),\ (\frac{3}{4},\frac{1}{4}),\ (\frac{1}{4},\frac{3}{4}),\ (\frac{3}{4},\frac{3}{4})$, following the procedure in Fig. \ref{Fig: dimer2} (a). We note that such deformation is adiabatic and should not spoil the higher-order topology since it only modifies the value of $\Delta_d$ to $-1$.     

In the Majorana representation, we can realize both the atomic and Wannier orbital configurations for $H_3$ by constructing a Majorana model shown in Fig. \ref{Fig: dimer2} (b). Notably, the unit cell now consists of four pairs of Majorana fermions (or equivalently, electron and hole pairs) at four inequivalent maximal Wyckoff positions. For convenience, we will use a two-component sublattice index ($i_x j_y$) with $i,j\in\{A,B\}$ to label the four maximal Wyckoff positions. For example, we refer ${\bf q}_{1a}$ as ($A_xA_y$) and ${\bf q}_{1b}$ as $(B_xA_y)$, which will be denoted as $(AA)$ and $(BA)$ for short.
As shown in Fig. \ref{Fig: dimer2} (b), there are three inequivalent Majorana bonds for $H_3$: (i) the strong green bonds $t$ that decide the positions of Wannier orbitals; (ii) the weak (solid and dashed) purple dimer bonds $m_y$ that remove the trivial edge modes on the $y$-edge; (iii) the weak (solid and dashed) red bonds $m_x$ that remove the trivial edge modes on the $x$-edge. Just from the schematic plot in Fig. \ref{Fig: dimer2} (a), we already expect that there will be one single unpaired Majorana mode (shown by the red and blue circles) on both the top-right and bottom-left corners of an inversion-symmetric finite-size geometry.

Following Fig. \ref{Fig: dimer2} (b), the corresponding Majorana Hamiltonian $H_3$ consists of three parts: $h_{3,g}$ for the green bonds, $h_{3,p}$ for the purple bonds, and the $h_{3,r}$ for the red bonds. In the Majorana representation, we have
\begin{eqnarray}
	h_{3,g} &=& i t \sum_{\bf R} [\beta_{{\bf R},AA} \alpha_{{\bf R},BB} + \beta_{{\bf R},AB} \alpha_{{\bf R}+{\bf a}_y,BA} \nonumber \\
	&& + \beta_{{\bf R},BA} \alpha_{{\bf R}+{\bf a}_x,AB} + \beta_{{\bf R},BB} \alpha_{{\bf R}+{\bf a}_d,AA}], \nonumber \\
	h_{3,p} &=& i m_x \sum_{\bf R} [\beta_{{\bf R},AA} \beta_{{\bf R},BA} + \alpha_{{\bf R},AA} \alpha_{{\bf R} - {\bf a}_x, BA} \nonumber \\
	&& + \beta_{{\bf R},AB} \beta_{{\bf R},BB} + \alpha_{{\bf R},AB} \alpha_{{\bf R} - {\bf a}_x, BB}], \nonumber \\
	h_{3,r} &=& i m_y \sum_{\bf R} [\beta_{{\bf R},AA} \beta_{{\bf R},AB} + \alpha_{{\bf R},AA} \alpha_{{\bf R} - {\bf a}_y, AB} \nonumber \\
	&& + \beta_{{\bf R},BA} \beta_{{\bf R},BB} + \alpha_{{\bf R},BA} \alpha_{{\bf R} - {\bf a}_y, BB}], \nonumber \\
	&&
\end{eqnarray}
where we have defined ${\bf a}_d = {\bf a_x + a_y}$ for simplicity. When transforming the Hamiltonian into momentum space, it is convenient to consider the Fourier transformations for the real-space Majorana operators. Take $\alpha$-type Majorana operator as an example,
\begin{eqnarray}
	\alpha_{{\bf R},ij} = \sum_{\bf k} e^{i{\bf k} \cdot ({\bf R} + {\bf r}_{ij} )} \alpha_{{\bf k},ij},
\end{eqnarray} 
where ${\bf r}_{ij}$ is the displacement of the sublattice $(ij)$ away from the unit-cell origin. The hermiticity of a Majorana operator requires
\begin{equation}
	\alpha_{{\bf k},ij} = \alpha_{-{\bf k},ij}^{\dagger}.
\end{equation}
We consider the following momentum-space Majorana basis
\begin{eqnarray}
	\Psi_M = && (\alpha_{{\bf k},AA}, \beta_{{\bf k},AA}, \alpha_{{\bf k},BA}, \beta_{{\bf k},BA}, \nonumber \\
	&&\  \alpha_{{\bf k},AB}, \beta_{{\bf k},AB}, \alpha_{{\bf k},BB}, \beta_{{\bf k},BB})^T,
\end{eqnarray}
under which the Hamiltonian is given by
\begin{widetext}
	\begin{equation}
	H_3 ({\bf k}) = \begin{pmatrix}
	0 & 0 & i m_x e^{i\frac{k_x}{2}} & 0 & i m_y e^{i\frac{k_y}{2}} & 0 & 0 & -i t e^{i\frac{k_x+k_y}{2}} \\
	0 & 0 & 0 & i m_x e^{-i\frac{k_x}{2}} & 0 & i m_y e^{-i\frac{k_y}{2}} & i t e^{-i\frac{k_x+k_y}{2}} & 0 \\
	-i m_x e^{-i\frac{k_x}{2}} & 0 &  0 & 0 & 0 & -i t e^{i\frac{k_x+k_y}{2}} & i m_y e^{i\frac{k_y}{2}} & 0 \\
	0 & -i m_x e^{i\frac{k_x}{2}} & 0 & 0 & i t e^{-i\frac{k_x+k_y}{2}} & 0 & 0 & i m_y e^{-i\frac{k_y}{2}} \\
	-i m_y e^{-i\frac{k_y}{2}} & 0 & 0 & -i t e^{i\frac{k_x+k_y}{2}} & 0 & 0 & i m_x e^{i\frac{k_x}{2}} & 0 \\
	0 &	-i m_y e^{i\frac{k_y}{2}} &  i t e^{-i\frac{k_x+k_y}{2}} & 0 & 0 & 0 & 0 & i m_x e^{-i\frac{k_x}{2}} \\
	0 & -i t e^{i\frac{k_x+k_y}{2}} & -i m_y e^{-i\frac{k_y}{2}} & 0 & -i m_y e^{-i\frac{k_x}{2}} & 0 & 0 & 0 \\
	i t e^{-i\frac{k_x+k_y}{2}} & 0 & 0 & -i m_y e^{i\frac{k_y}{2}} & 0 & -i m_x e^{i\frac{k_x}{2}} & 0 & 0 \\
	\end{pmatrix}
	\end{equation}
\end{widetext}

Interestingly, $H_3$ can be diagonalized analytically and we find that the energy eigenvalues are completely ${\bf k}$-independent:
\begin{equation}
	E=\pm \sqrt{t^2+(m_x\pm m_y)^2}.
\end{equation}

To further clarify the higher-order nature of $H_3$, we consider an inversion-symmetric open geometry for $H_3$ and calculate the energy spectrum with $t=1$ and $m_x=m_y=0.5$. As shown in Fig. \ref{Fig: dimer2} (c) and its inset, we confirm the existence of a pair of corner-localized Majorana zero modes, which agrees with the prediction from the counting rule.  

\subsubsection{Diagnosis from ${\bf k}$-Space Symmetry Indicators} \label{Subsec: Beyond indicator}

We also calculate the parity data for the higher-order TSC phase of $H_3({\bf k})$, which is shown in Fig. \ref{Fig: dimer2} (d). The inversion operator is given by
\begin{equation}
	{\cal I} = \tau_z\otimes \begin{pmatrix}
	1 & & & \\
	& e^{i k_x} & & \\
	& & e^{i k_y} & \\
	& & & e^{i(k_x+k_y)} \\
	\end{pmatrix}
   \label{eq:inversion for H3}
\end{equation}
in the fermion basis 
\begin{eqnarray}
	\Phi_F &=& (c_{{\bf k},AA}, c_{{\bf k},BA}, c_{{\bf k},AB}, c_{{\bf k},BB}, \nonumber \\
	&& c_{-{\bf k},AA}^{\dagger}, c_{-{\bf k},BA}^{\dagger}, c_{-{\bf k},AB}^{\dagger}, c_{-{\bf k},BB}^{\dagger})^T.
\end{eqnarray}
Interestingly, $H_3$ has an equal number of ``$+$" and ``$-$" parity eigenvalues for every high-symmetry points. To understand this parity data, we note that $m_x$ and $m_y$ are essentially dimer bonds that lead to unit-cell enlargement. Therefore, the BZ with finite $m_{x,y}$ is actually folded in both $k_x$ and $k_y$ directions, in comparison to the original BZ without the dimers. Similar to the double-stacking model $H_2({\bf k})$, the inversion eigenstates for the new high-symmetry momenta after the BZ folding are essentially the bonding and anti-bonding combinations of energy eigenstates at these momenta that transform from one to another under inversion. As a result, we will always have an equal number of energy eigenstates with ``$+$" and ``$-$" parity eigenvalues at all high-symmetry momenta. While $H_3$ appears to be ``featureless" in terms of its parity data, Ref.~\cite{huang2021indicator} has demonstrated the applicability of symmetry indicator theory to understand its topology, which we will briefly describe below. 
	
Since ${\cal I}$ is ${\bf k}$-dependent, it is crucial to introduce a flat-band BdG Hamiltonian $H_\text{ref} = \tau_z \otimes \mathbb{1}_4$, which exhibits the following parity data according to Eq.~\ref{eq:inversion for H3},
\begin{eqnarray}
 	&&[(-,-,-,-)_{\Gamma}, (-,+,-,+)_X, \nonumber \\
 	&& (-,-,+,+)_Y, (-,+,+,-)_M ].
\end{eqnarray}
Define $n_{K_i}$ as the number of positive-parity states at $K_i$. The $\mathbb{Z}_4$ inversion indicator is defined by
\begin{equation}
	\kappa = \sum_{K_i} [n_{K_i}(H_3) - n_{K_i}(H_\text{ref})] = 2,
\end{equation}
which directly implies a 2nd-order topology. This analysis agrees with our real-space counting rule and the numerical corner-geometry simulation.

\section{Proposal of a Real-Space Topological Diagnosis} \label{Sec: Conjecture}

\begin{figure}[t]
	\centering
	\includegraphics[width=0.5\textwidth]{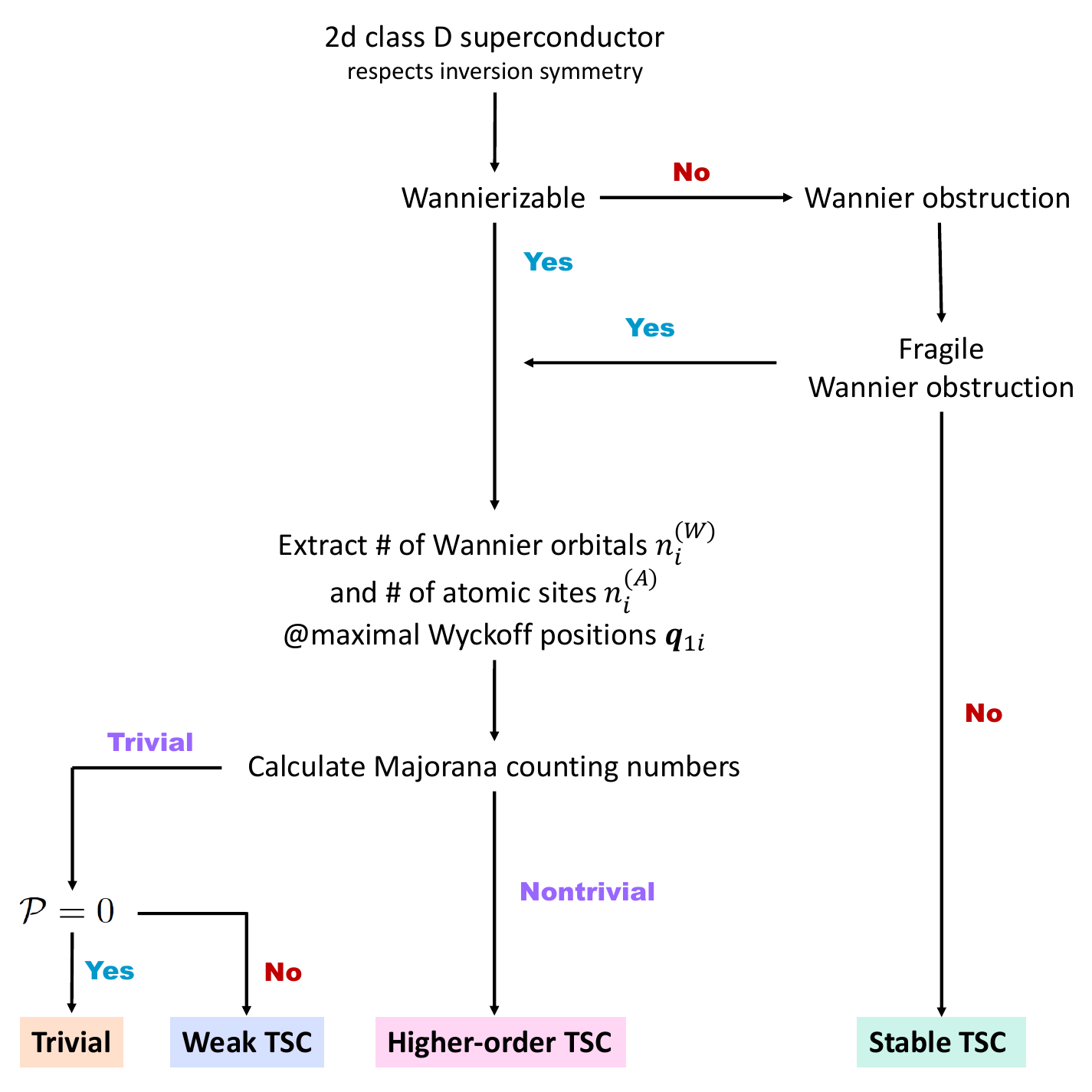}
	\caption{The flow diagram of our proposed real-space diagnosis for 2d class-D inversion-protected higher-order TSCs.}
	\label{Fig: Diagnose}
\end{figure}

Although the Majorana counting rule is derived in the Kitaev limit, the resulting counting procedure only involves the position information of BdG Wannier orbitals. It is thus natural to expect that the counting rule should also hold for general Wannierizable superconductors, including those that might not manfiest a Kitaev-limit description. In this case, we expect that a rigorous proof of the real-space counting rules for all Wannierizable systems could be possible, but challenging to achieve, while the Kitaev limit offers a special shortcut to access these relations. With this conjectured applicability of counting rule in mind, we propose in this section a real-space diagnosis for higher-order topology for 2d class D systems with inversion symmetry, as schematically shown in Fig.~\ref{Fig: Diagnose}. 

In the following, we briefly go through the proposed procedure for this conjectured universal diagnosis:
\begin{itemize}
	\item {\bf Step 1}: Given a 2d class D superconductor with inversion symmetry, first check if its Wannierizability. If the Wannier obstruction is fragile, remove it. If not, characterize the system as a stable topological superconductor, where the term ``stable" is defined to describe the robustness of Wannier obstruction.
	\item {\bf Step 2}: If there is no stable Wannier obstruction, extract the numbers of Wannier orbitals and atomic sites at each maximal Wyckoff positions. Calculate Majorana counting numbers.
	\item {\bf Step 3}: If the counting numbers are non-trivial, the system is a higher-order TSC. If not, the system could be either a trivial superconductor or a weak TSC, depending on the explicit value of polarization ${\cal P}$.     
\end{itemize}

To demonstrate our conjectured universal diagnosis, in the next section, we will discuss a known higher-order TSC model in the literature. We will first show that this model actually hosts fragile Wannier obstruction and is definitely beyond the Kitaev limit. Nevertheless, we will follow the universal diagnosis to first remove the Wannier obstruction and connect the new Wannierizable system to a well-defined Kitaev superconductor. This allows us to understand the origin of higher-order topology for the original model through a very simple Majorana counting, which verifies our diagnosis.   

\section{Application to a Realistic Higher-order Topological Superconductor}\label{Sec: Fragile}

Let us now consider a realistic higher-order TSC that has been discussed in the literatures \cite{khalaf2018higher,hsu2019inversion}. This model consists of one $p+ip$ chiral TSC and another $p-ip$ chiral TSC \cite{khalaf2018higher}, and is thus a superconducting version of the ``shift insulator" in Ref. \cite{liu2019shift}. We first present a minimal tight-binding model of such system and explicitly show the existence of Majorana corner modes by deriving an effective edge theory analytically. We then calculate its Wilson loop spectrum to confirm the inversion-protected Wannier obstruction, manifested in the nontrivial Wilson loop winding. However, such Wilson loop winding can be trivialized if we couple the system with additional Wannier orbitals, which thus confirms the fragility of Wannier obstruction \cite{fragile}. By matching the stacked chiral TSC model with a superposition of Kitaev building blocks, we provide an understanding of its higher-order topology with the help of our Majorana counting rule.

\subsection{Model Hamiltonian for Stacked Chiral TSCs}

The minimal tight-binding model for two decoupled chiral TSC is given by
\begin{eqnarray}
H_4^{(0)} &=& [m_0 + m_1 (\cos k_x + \cos k_y)] \Gamma_5 \nonumber \\
&& + v (\sin k_x \Gamma_1 + \sin k_y \Gamma_2)  
\end{eqnarray}
where we have defined the generator of $4\times 4$ $\Gamma$ matrices as
\begin{eqnarray}
	\Gamma_1 &=& \tau_x \otimes \sigma_0,\ \ \Gamma_2 = \tau_y \otimes \sigma_0,\ \ \Gamma_3 = \tau_z \otimes \sigma_x,\nonumber \\
	 \Gamma_4 &=& \tau_z \otimes \sigma_y,\ \ \Gamma_5 = \tau_z \otimes \sigma_z,
\end{eqnarray}
These $\Gamma$ matrices satisfy an anti-commutation relation $\{\Gamma_i, \Gamma_j \} = 2\delta_{ij}$ for $i,j\in\{1,2,3,4,5\}$. The other ten $\Gamma$ matrices can be generated by $\Gamma_{jk} = \frac{1}{2i}[\Gamma_j, \Gamma_k]$ for $j\neq k$. 

Note that $H_4^{(0)}$ is block-diagonal and can be written as a direct sum of a $p+ip$ chiral TSC $h_+$ and another $p-ip$ chiral TSC $h_-$. In particular,  
\begin{eqnarray}
	h_{\pm} ({\bf k}) &=& \pm [m_0 + m_1 (\cos k_x + \cos k_y)] \tau_z \nonumber \\
	&& + v (\sin k_x \tau_x + \sin k_y \tau_y).
\end{eqnarray}
We have defined the particle-hole symmetry as $\Xi = \Gamma_1 {\cal K}$ and the inversion symmetry as ${\cal I}=\Gamma_5$, where ${\cal K}$ is the complex conjugation. 

We now consider adding symmetry-allowed perturbation $H_4^{(1)}$ to remove the accidental edge modes of $H_4^{(0)}$. To be specific, we hope to find a constant matrix $A$ to preserve both inversion ${\cal I}$ and the particle-hole symmetry $\Xi$. Namely, $A$ should satisfy (i) $\{A, \Xi\} = 0$; (ii) $[A,{\cal I}]=0$. We find the following choice of $A$:
\begin{equation}
	A = \{\Gamma_5, \Gamma_{12}, \Gamma_{14}, \Gamma_{24}\},
\end{equation}
which inspires us to define
\begin{equation}
	H_4^{(1)} = g_1 \Gamma_{14} + g_2 \Gamma_{24} + g_3 \Gamma_{12},
\end{equation}
where we have ignored $\Gamma_5$ in $H_4^{(1)}$ since it is already contained in $H_4^{(1)}$. The complete Hamiltonian for stacked chiral TSC is thus $H_4 = H_4^{(0)} + H_4^{(1)}$.

\subsection{Higher-order Topology from a Boundary Perspective}
Before providing any numerical results for $H_4$, we first demonstrate the origin of its higher-order topology from an analytical boundary perspective. We will derive an effective analytical boundary theory for $H_4$ in a disk geometry and explicitly show the existence of corner-localzied Majorana zero modes, similar to the approaches in Ref. \cite{hsu2019inversion,zhang2019dirac}. 
  
We first treat $H_4^{(1)}$ as a small perturbation and expand $h_{\pm}$ around $\Gamma$ point. This leads to an effective Hamiltonian around $\Gamma$ for both chiral TSC block,
\begin{equation}
h_{\pm}^{\Gamma} ({\bf k}) = \pm[\tilde{m}_0 - \tilde{m}_1 (k_x^2 + k_y^2)] \tau_z + v (k_x \tau_x + k_y \tau_y),
\end{equation}
where we have defined $\tilde{m}_0 = (m_0 + 2m_1)$ and $\tilde{m}_1 = \frac{m_1}{2}$. In the polar coordinate $r=\sqrt{x^2+y^2}$ and $\theta = \tan^{-1} \frac{y}{x}$, we have $k_{\pm} = e^{\pm i\theta} ( k_r \pm ik_{\theta})$ with $k_r = -\partial_r$ and $k_{\theta} = -\frac{i}{r}\partial_{\theta}$. Up to ${\cal O}(k)$, $h_{\pm}^{\Gamma}$ can be written as 
\begin{equation}
h_{\pm}^{\Gamma} (r,\theta) = 
\begin{pmatrix}
\pm \tilde{m}_0 & v e^{-i\theta} (-i\partial_r -\frac{1}{r}\partial_{\theta}) \\
v e^{i\theta} (-i\partial_r + \frac{1}{r}\partial_{\theta}) & \mp \tilde{m}_0 \\
\end{pmatrix}.
\end{equation}
We consider a disk geometry with a radius $R$ and solve for the Majorana wavefunction that is exponentially localized at $r=R$. In the large $R$ limit, we find a single Majorana solution $\psi_{\pm}$ for $h_{\pm}$: 
\begin{equation}
	\psi_{\pm} (l,r,\theta) = {\cal N}_{\psi} e^{\pm \frac{m}{v}(r-R)} e^{il\theta} \begin{pmatrix}
	 e^{-\frac{i}{2}(\theta \pm \frac{\pi}{2})} \\
	 e^{\frac{i}{2}(\theta \pm \frac{\pi}{2})} \\
	\end{pmatrix}
\end{equation}
with $l\in \mathbb{Z}$ and a normalization factor ${\cal N}_{\psi}$. The energy dispersion for $\psi_{\pm}(l,r,\theta)$ is 
\begin{equation}
	E_{\pm,l} = \pm \frac{vl}{r}.
\end{equation}  
Therefore, $\psi_{+}$ and $\psi_{-}$ represent a pair of chiral Majorana edge modes propagating in the opposite directions. Now we are ready to project the perturbation $H_4^{(1)}$ onto the chiral Majorana basis and we arrive at an effective edge Hamiltonian
\begin{equation}
	H_\text{edge} = \frac{vl}{r} \sigma_z - (g_1 \sin\theta - g_2 \cos\theta) \sigma_y.
\end{equation}
Therefore, the edge spectrum is given by
\begin{equation}
	E_\text{edge} = \pm \sqrt{(\frac{vl}{r})^2 + (g_1^2+g_2^2) \sin (\theta-\theta_0)},
\end{equation}
where we have defined $\theta_0 = \tan^{-1} \frac{g_2}{g_1}$. Clearly, the boundary gap closes for the chiral Majorana modes with $l=0$ only when $\theta=\theta_0$ and $\theta=\theta_0+\pi$.
Therefore, the system hosts a pair of corner-localized Majorana zero modes at these two inversion-related angles and is thus higher-order topological by definition.

\subsection{Fragile Wannier Obstruction}

\begin{figure}[t]
	\centering
	\includegraphics[width=0.5\textwidth]{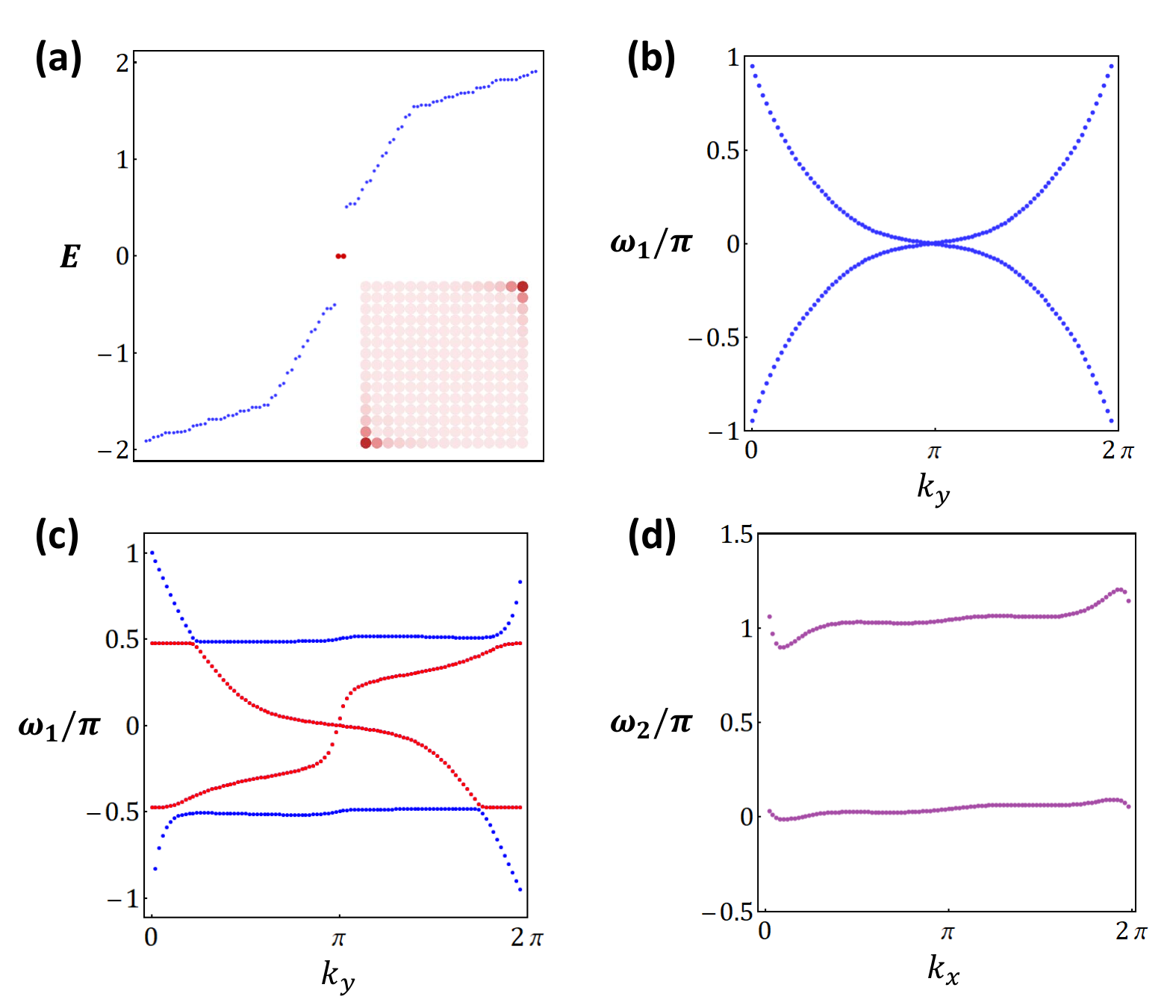}
	\caption{We numerically verify in (a) and the inset that the stacked chiral TSC model $H_4$ is higher-order topological by calculating its energy spectrum on an open geometry and confirm its corner-localized Majorana zero modes. In (b), we calculate the 1d $x$-directed Wilson loop spectrum $\omega_1(k_y)$. $\omega_1$ is found to host a nontrivial winding pattern which directly implies the existence of Wannier obstruction. In (c), we explicitly demonstrate the fragility of Wannier obstruction by coupling $H_4$ to a Wannierizable system $H_w$ and thus ``unwind" the Wilson loop pattern. In (d), we calculate the $y$-directed nested Wilson loop $\omega_2(k_x)$ for the red Wannier band sector in (c).}
	\label{Fig: Fragile}
\end{figure}

We now consider placing $H_4$ on a $15\times 15$ lattice and calculate its energy spectrum. As shown in Fig. \ref{Fig: Fragile} (a) and its inset, there exists a pair of corner-localized Majorana zero modes, which establishes $H_4$ as a higher-order TSC. In particular, the model parameters are chosen to be $m_0 = -m_1 = 3$ and $g_1=g_2 = 0.5$. According to our boundary theory, the corner Majorana modes are predicted to be localized on the inversion-related boundary positions, which are characterized by the polar angle $\theta = \theta_0 = \tan^{-1} 1 = \frac{\pi}{4}$ and $\theta=\frac{5\pi}{4}$. This agrees well with our numerical findings shown in the inset of Fig. \ref{Fig: Fragile} (a). \\

The Wannier obstruction of $H_4$ is clearly revealed by calculating the bulk Wilson loop spectrum $\omega_1^x(k_y)$, which is often known as the ``Wannier band spectrum" \cite{benalcazar2017quantized}. As shown in Fig. \ref{Fig: Fragile} (b), the Wilson loop $\omega_1^x(k_y)$ exhibits a nontrivial winding pattern which prohibits a symmetric and localized Wannier representation for $H_4$. Such winding pattern is relatively robust by itself due to the inversion-symmetry protection \cite{alexandradinata2014wilson}.  

However, if we couple $H_4$ with another Wannierizable system $H_w$, we can unwind the Wilson loop (gap out the Wannier bands) to recover the Wannierizability \cite{po2018fragile,wieder2018axion}. To achieve this, we consider a composite system $H_5$, with
\begin{equation}
	H_5 = \begin{pmatrix}
	H_4 & h_c \\
	h_c^{\dagger} & H_w
	\end{pmatrix}.
\end{equation}
We have constructed $H_w$ by placing one Wannier orbital at ${\bf q}_1 = (\frac{1}{4}, \frac{1}{4})$ and another Wannier orbital at ${\bf q}_2 = (-\frac{1}{4}, -\frac{1}{4})$. Note that one can symmetrically move both Wannier orbitals to ${\bf q}_{1a}$ without spoiling the adiabacity. Therefore, $H_w$ is adiabatically equivalent to stacking two building blocks $\kappa_a$ at ${\bf q}_{1a}$, which is both Wannierizable and topologically trivial.

Specifically, we consider the BdG basis $\Phi_w = (c_{1,{\bf k}},c_{1,-{\bf k}}^{\dagger},c_{2,{\bf k}},c_{2,-{\bf k}}^{\dagger})^T$ for $H_w$, where $c_{i,{\bf k}}$ annihilates an electron at ${\bf q}_i$ for $i=1,2$. Then $H_w$ and the coupling matrix $h_c$ are given by

\begin{eqnarray}
	H_w &=& t \begin{pmatrix}
	\epsilon_0 & 0 & f({\bf k})^2 & 0 \\
	0 & -\epsilon_0 & 0 & -f({\bf k})^2 \\
	[f({\bf k})^*]^2 & 0 & \epsilon_0 & 0 \\
	0 & -[f({\bf k})^*]^2 & 0 & -\epsilon_0 \\
	\end{pmatrix}, \nonumber \\
	h_c &=& g_c \begin{pmatrix}
	f({\bf k})^* & 0 & f({\bf k}) & 0 \\
	0 & -f({\bf k})^* & 0 & -f({\bf k}) \\
	0 & 0 & 0 & 0 \\
	0 & 0 & 0 & 0 \\
	\end{pmatrix}.
\end{eqnarray}
where we have defined $f({\bf k}) = e^{\frac{i}{4}(k_x+k_y)}$. 

For our purpose, we choose a large $\epsilon_0=30$ such that the BdG bands of $H_w$ stay away from the Fermi level. We also choose $t=0.5$ and $g_c=4$ and calculate the $x$-directed Wilson loop for the composite system $H_5$. As shown in Fig. \ref{Fig: Fragile} (c), now the Wannier bands are gapped out and can be separated into two disjoint segments, which are plotted in red and blue, respectively. This unwinding pattern directly suggests that the previous Wannier obstruction for $H_4$ is indeed removed, which is a hallmark for fragile Wannier obstruction.

\subsection{Nested Wilson Loop, Parity Data, and Majorana Counting}

\begin{table}[t]
	\includegraphics[width=0.5\textwidth]{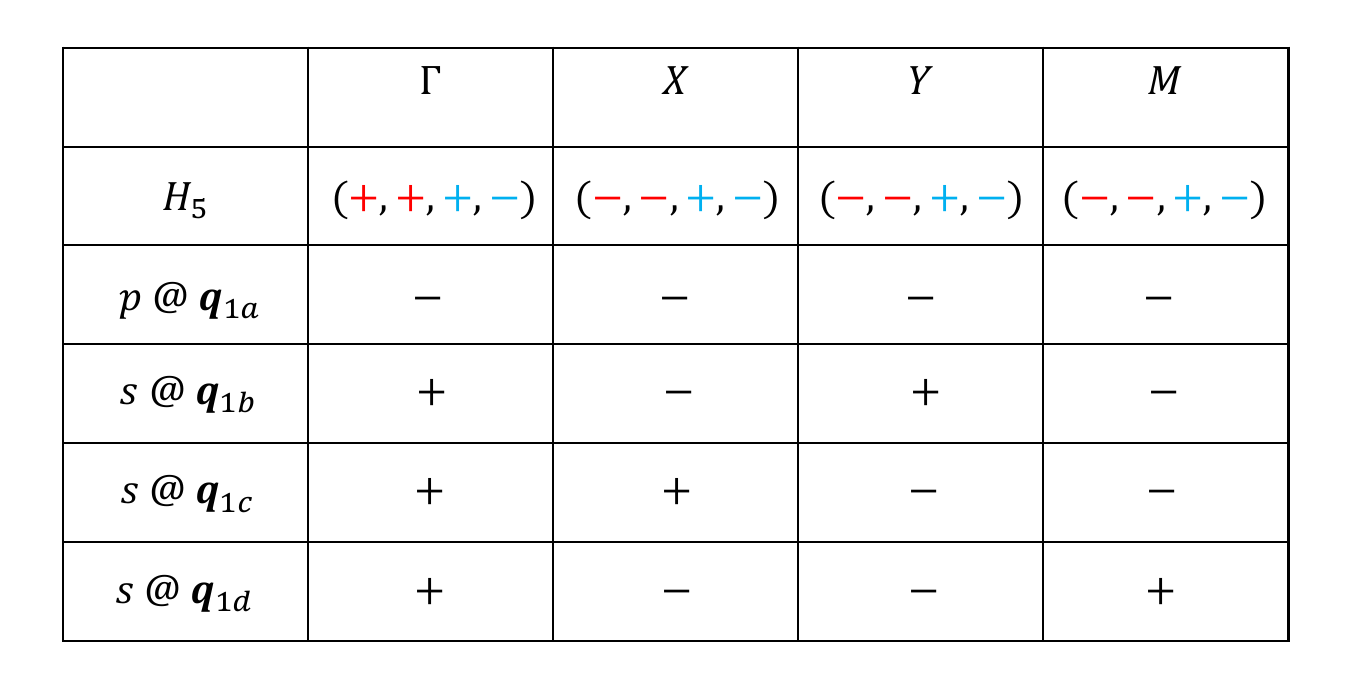}
	\caption{This table shows the decomposition of the composite system $H_5$ with respect to Kitaev building blocks, from the perspective of parity data. For the parity data of $H_5$, the data in red (blue) color shows the contribution from $H_4$ ($H_w$). About the notation of the decomposition, for example, we have used ``$p@{\bf q}_{1a}$" to denote a Kitaev building block with a $p$-like orbital sitting at ${\bf q}_{1a}$.}
	\label{Table: Parity for H5}
\end{table}  

To perform our universal diagnosis, note that it is generally challenging to find an adiabatic path to connect the Wannierizable composite system $H_5$ and a corressponding Kitaev superconductor. Consequently, we will try to establish such adiabatic connection in an indirect way. We will explicitly show that both its (nested) Wilson loops \cite{benalcazar2017quantized} and the parity data indicate a unique decomposition of $H_4$ in terms of the Kitaev building blocks. This decomposition allows us to explain the higher-order topological origin of the stacked chiral TSC model $H_4$ with our Majorana counting rule. 

Physically, the value of Wilson loops $\omega_1^x(k_y)$ is the expectation value of $x$-position operator $\hat{x}$ in the Wannier representation. Therefore, a gapped Wannier band spectrum in Fig. \ref{Fig: Fragile} (c) directly implies the existence of two spatially separated electron clouds along $x$-direction within the unit cell. To be specific, the red (blue) Wannier bands correspond to an electron cloud localized around $x=0$ ($x=\frac{1}{2}$). Please note that the pseudo-energy unit of Wannier bands and that of spatial coordinates differ by a factor of $2\pi$. 

To further extract the position information for the electron clouds along $y$-direction, we calculate the nested Wilson loop $\omega_2^y(k_x)$ for each colored group of Wannier bands. As shown in Fig.~\ref{Fig: Fragile} (d), we find two nested Wannier bands localized around $\omega_2^y = 0$ and $\omega_2^y = \frac{1}{2}$, respectively. Physically, this implies that the red electron cloud at $x=0$ can be further divided into two separated smaller electron clouds along $y$-direction. In particular, one cloud sits around ${\bf q}_{1a} = (0,0)$ and the other locates at ${\bf q}_{1c} = (0,\frac{1}{2})$. A similar nested Wilson loop calculation can be performed for the blue Wannier band sector, which shows a similar spectrum to Fig.~\ref{Fig: Fragile} (d). This suggests the existence of two electron clouds at ${\bf q}_{1b}$ and ${\bf q}_{1d}$, respectively. 

However, one should be careful about interpreting the (nested) Wilson loop results as the actual positions of Wannier orbitals in real-space. Since the (nested) Wilson loops are essentially projected position operators for a specific (Wannier) band sector, the projected $\hat{x}$ and $\hat{y}$ operators do not generally commute. For example, we could calculate the $y$-directed Wilson loop $\omega_1^y$ and the $x$-directed nested Wilson loop $\omega_2^x$, instead of the $\omega_1^x$ and $\omega_2^y$ that we have calculated. Then it is possible that $(\omega_1^x,\omega_2^y) \neq (\omega_1^y,\omega_2^x)$. While this ambiguity does not exist for our present model, an example with this issue does exist and was addressed in Ref. \cite{benalcazar2017quantized}.

To further rule out the ambiguity of the Wilson loop calculations, we further calculate the parity data for the composite system $H_5$, as shown in Table \ref{Table: Parity for H5}. In particular, we can trust the parity data for $H_5$ simply because it does have momentum-independent inversion representation. To demonstrate, we have highlighted the parity contribution for $H_4$ in red and that for the Wannierizable system $H_w$ in blue. We note that the parity data for $H_5$ is the same as that for a face-to-face stacking of one p-like orbital at ${\bf q}_{1a}$ and three s-like orbital at the other three maximal Wyckoff positions, as shown in Table \ref{Table: Parity for H5}. It is easy to see that such parity data decomposition is indeed {\it unique}. Then the parity data and the Wilson loop calculations reach a consistent decomposition relation for $H_5$ that
\begin{equation}
	H_5 \equiv  \kappa_a (p) \oplus \kappa_b (s) \oplus \kappa_c (s) \oplus \kappa_d (s),
\end{equation}
where we have denote a $\kappa_i$ with an $\alpha$-type orbital as $\kappa_i(\alpha)$. Here ``$\equiv$" denotes the adiabatic equivalence relation between the two systems and ``$\oplus$" denotes the stacking operation of Kitaev building blocks.

Recall that $H_w \equiv \kappa_a (s) \oplus \kappa_a (p)$. Then we have
\begin{equation}
	H_4 \equiv  H_5 \ominus H_w \equiv \kappa_b (s) \oplus \kappa_c (s) \oplus \kappa_d (s) \ominus \kappa_a (s).
\end{equation}
Namely, $H_4$ is equivalent to a stacking of $\kappa_{b,c,d}$ with an additional ``subtraction $\ominus$" of $\kappa_a$. From the Majorana counting rule, we have shown that the face-to-face stacking of $\kappa_{b,c,d}$ is higher-order topological but $\kappa_a$ is not. As a result, the stacked chiral TSC model $H_4$ is then naturally higher-order topological, which clearly verifies our diagnosis.

\section{Conclusions and Discussions} \label{Sec: Conclusion}

To conclude, we have proposed both Kitaev limit and its building blocks as a new approach for constructing and diagnosing 2d higher-order topological superconductors with inversion symmetry. Different from the traditional ${\bf k}$-space classification schemes, our approach starts from a complete set of real-space Kitaev building blocks, based on which we derive a simple real-space counting rule for diagnosing corner Majorana modes. The Majorana counting rule is expected to work for general 2d higher-order TSCs within this symmetry class, with which we arrive at simple recipes for constructing various paradigmatic higher-order TSC models by stacking the building blocks. We also demonstrate an application of the real-space counting rule to a non-Kitaev-limit model that carries a fragile Wannier obstruction.

Recently, there is an interesting proposal about using ``pairing obstruction" instead of Wannier obstruction to understand the topological behavior of superconductors \cite{schindler2020pairing}. Following the original definition in Ref. \cite{read2000paired}, the pairing obstruction is characterized by a real-space two-point function $g_{\bf r r'}$ that describes the spatial profile of the Cooper pairs. In particular, a BdG system is pairing obstructed and thus proposed to be topological if $g_{\bf r r'}$ falls off as a polynomial function of $|{\bf r-r'}|$. In other words, the pairing obstruction occurs if the Fourier transform of $g_{\bf r r'}$ is singular. We note that our Kitaev building blocks $\kappa_{b,c,d}$ do have similar pairing obstructions because of the inter-atomic-site Majorana bonds. Therefore, it is easy to check that all of the Kitaev superconductors with higher-order topology in our work are indeed pairing obstructed.

Generalizations of our building block construction to superconductors with other crystalline or internal group symmetries are possible. For example, in a follow-up work by two of the authors, we extend our framework to 2d class DIII systems with a two-fold rotation symmetry $C_2$~\cite{vu2020C2}. The specialty of this symmetry class lies in the universally equivalent $C_2$ symmetry data at every high-symmetry momentum, which is independent of the underlying band topology. Systems in this class thus cannot be symmetry indicated~\cite{chen2022topo}, while our real-space diagnosis has been proved powerful to analyze the corner Majorana physics in this case. As a future direction, it would be interesting to understand the relationship between our counting rule and the symmetry indicators. 

Our approach is also ready to be extended for 3d higher-order TSCs. Note that in 3d class D systems, inversion symmetry can support 2nd and 3rd order topologies. The latter case features corner Majorana modes similar to the 2nd-order phase in 2d, which would be either Wannierizable or fragilely Wannier obstructed. To derive a similar counting rule in 3d, the increase of spatial dimensions is expected to lead to eight inequivalent Kitaev building blocks. Besides, a non-zero polarization could now lead to either surface or hinge-localized dangling BdG Wannier orbitals, depending on its spatial orientation. All of the above details will need to be handled carefully to arrive at the correct Majorana counting rule, which is complicated but doable. Meanwhile, 3d 2nd-order TSC phases are usually not Wannierizable, which manifests in their dispersing hinge modes. However, they generally yield a dimensional-reduction picture that, either $k_z=0$ or $k_z=\pi$ plane of the 3d system is topologically equivalent to a 2d 2nd-order TSC. In this case, we can always apply our 2d real-space diagnosis to understand the Majorana corner modes of the ${\bf k}$-space 2d subsystem, further offering an interpretation to the Majorana hinge modes. We leave a detailed discussion on this subject to future works.

\begin{acknowledgments}
RXZ is grateful to DinhDuy Vu, Jiabin Yu, and especially Yi-Ting Hsu and Sheng-Jie Huang for helpful discussions. We thank the anonymous referee for the helpful comments on the symmetry indicator theory. This work is supported by the Laboratory for Physical Sciences and Microsoft. RXZ acknowledges a JQI Postdoctoral Fellowship at the University of Maryland and a start-up funding at the University of Tennessee. JS was supported by the NSF-DMR1555135 (CAREER). This research was supported in part (through helpful discussions at KITP) by the National Science Foundation under Grant No. NSF PHY-1748958.
\end{acknowledgments}

\appendix

\section{Momentum-dependence of Symmetry Representation} \label{Appendix A}
In Sec. \ref{Sec: Atomic Sites}, we have discussed that when the unit cell of a lattice model is inversion-breaking, its matrix representation of inversion symmetry in momentum space is necessarily momentum-dependent. In this appendix, we will discuss two explicit examples to demonstrate the deep connection between an inverion-breaking unit cell in real space and the symmetry matrix representation in momentum space. 

\subsection{Staggered Model in One Dimension} 

\begin{figure}[t]
	\includegraphics[width=0.45\textwidth]{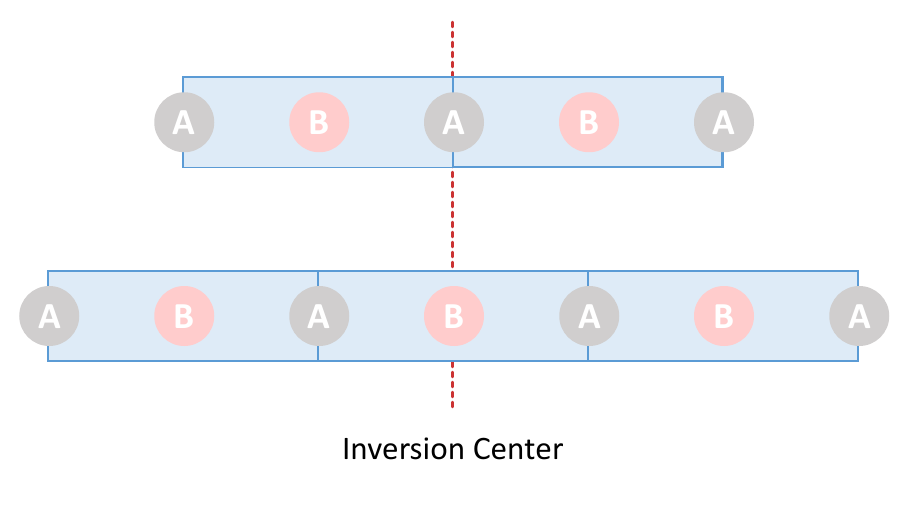}
	\caption{An inversion-symmetric open boundary for the 1d stagger model always contains a fractional unit cell on the boundary, where one atom $B$ is necessarily removed. The inversion center could locate at either atom $A$ or atom $B$.}
	\label{Fig: Stagger model}
\end{figure}

Let us first consider a 1d inversion-symmetric lattice system with two types of atoms $A$ and $B$ within one unit cell, as shown in Fig. \ref{Fig: Stagger model}. We choose the unit cell convention such that atom A locates at the unit-cell origin with $r_A = 0$ and atom B locates at the center of the unit cell with $r_B=1/2$. In particular, we turn on an on-site potential $\mu_A$ for atom $A$ as well as an $\mu_B$ atom $B$, and further require the potential distribution to be staggered with $\mu_A \neq \mu_B$. As shown in Fig. \ref{Fig: Stagger model}, the inversion center will then coincide with either atom $A$ or atom $B$ for an open boundary system, which depends on the oddness of the number of unit cells. Since $A$ and $B$ locate at two distinct maximal Wyckoff positions for the 1d unit cell, the unit cell inevitably breaks inversion symmetry. Therefore, we expect that the matrix representation of inversion symmetry will necessarily depend on crystal momentum $k$. 

To verify this, we construct a real-space tight-binding Hamiltonian for this 1d staggered model as
\begin{eqnarray}
	H_\text{stagger} &=& \sum_{R} \sum_{i=A,B} \frac{\mu_i}{2} c^{\dagger}_{i,R} c_{i,R}  \nonumber \\
	&+&  t \sum_{R} [c^{\dagger}_{A,R} c_{B,R} + c^{\dagger}_{A,R} c_{B,R-1}] + h.c.
\end{eqnarray}
We perform Fourier transformation and arrive at the following momentum-space Hamiltonian 
\begin{equation}
	H_\text{stagger}(k) = \begin{pmatrix}
	\mu_A & t(1+e^{-i k}) \\
	t(1+e^{i k}) & \mu_B \\
	\end{pmatrix}.
\end{equation}      
The basis function for this Hamiltonian is given by
\begin{equation}
	\Phi = (c_{A,k},c_{B,k})^T.
\end{equation}
Since the inversion operation ${\cal I}$ leaves both atoms $A$ and $B$ invariant, the most general form for the matrix representation of ${\cal I}$ is given by
\begin{equation}
	{\cal I} = \begin{pmatrix}
	a_1(k) & \\
	& a_2(k) \\
	\end{pmatrix},
\end{equation}
where $a_{1,2}(k)$ are general complex functions of $k$. By imposing
\begin{equation}
	{\cal I} H_\text{stagger} {\cal I}^{\dagger} = H_\text{stagger} (-k),
\end{equation}
we arrive at the following constraints
\begin{equation}
	|a_{1,2}(k)|^2=1,\ \ a_1({\bf k}) a_2({\bf k})^* = e^{i k}.
\end{equation}
Therefore, we find that
\begin{equation}
{\cal I} = a_1(k) \begin{pmatrix}
1 & \\
& e^{-ik} \\
\end{pmatrix},
\end{equation}
where $a_1(k)$ can be treated as an unimportant $U(1)$ phase factor. 

On the other hand, if we consider the following modified basis,
\begin{equation}
	\tilde{\Psi} = (c_{A,k}, e^{-ik/2} c_{B,k})^T.
\end{equation}
Then the inversion operator is simply ${\cal I} = \mathbb{1}_2$ and $k$-independent. But the Hamiltonian now becomes
\begin{eqnarray}
	\tilde{H}_\text{stagger} = \begin{pmatrix}
	\mu_A & 2t\cos \frac{k}{2} \\
	2t\cos \frac{k}{2} & \mu_B \\
	\end{pmatrix},
\end{eqnarray}
which is clearly $4\pi$-periodic. This agrees with our expectation that if the unit cell breaks inversion and $H_\text{stagger}(k)$ is $2\pi$-periodic, the representation of ${\cal I}$ must explicitly depends on $k$. \\

\subsection{Inversion Symmetry of Double-stacking Model $H_2({\bf k})$}

Now let us construct the inversion operation for the minimal double-stacking model $H_2({\bf k})$.

If we follow the Fourier transformation in Eq. \ref{Eq: Fourier transform} and the Hamiltonian in Eq. \ref{Eq: H2(k)}, the inversion operation is simply ${\cal I} = \tau_z \otimes \sigma_0$. But we should notice that the Hamiltonian in Eq. \ref{Eq: H2(k)} is indeed $4\pi$-periodic instead of $2\pi$-periodic. 

Instead, we can follow the basis convention in Sec. \ref{Subsec: HOTSC in H2}, where the Fourier transform are defined with respect to lattice unit cells. In this convention, the Hamiltonian is guaranteed to be $2\pi$-periodic. Then the general form for inversion operation is given by
\begin{equation}
	{\cal I} = \begin{pmatrix}
	a_1({\bf k}) & & & \\
	& a_2({\bf k}) & & \\
	& & a_3({\bf k}) & \\
	& & & a_4({\bf k}) \\
	\end{pmatrix},
\end{equation}
where $a_i({\bf k})$ are assumed to be unitary complex functions of ${\bf k}$ with $|a_i({\bf k})|^2=1$. Since we require
\begin{equation}
{\cal I} H_2({\bf k}) {\cal I}^{\dagger} = H_2({\bf k}) (-k),
\end{equation}
we arrive at the following constraints
\begin{eqnarray}
	a_1({\bf k}) a_2({\bf k})^* &=& e^{-ik_y},\ \ a_1({\bf k})a_3({\bf k})^* = -1, \nonumber \\
	a_2({\bf k})a_3({\bf k})^* &=& -e^{i k_y},\ \ a_1({\bf k}) a_4({\bf k})^* = -e^{-ik_y},  \nonumber \\
	a_3({\bf k})a_4({\bf k})^* &=& e^{-i k_y},\ \ a_2({\bf k}) a_4({\bf k})^* = -1. \nonumber \\
\end{eqnarray}
Therefore, the inversion operator for $H_2({\bf k})$ is
\begin{equation}
	{\cal I} = a_1({\bf k}) \begin{pmatrix}
	1 & & & \\
	& e^{i k_y} & & \\
	& & -1 & \\
	& & & -e^{ik_y} \\
	\end{pmatrix},
\end{equation}
which is exactly the one shown in Sec. \ref{Sec: displaced stacking H2} up to an unimportant $U(1)$ phase factor $a_1({\bf k})$.

\bibliography{BuildingBlocks}

\begin{thebibliography}{84}
\bibitem [{hig()}]{highertopo}%
  \BibitemOpen
  \href@noop {} {}\bibinfo {note} {We note that there are multiple definitions
  of ``higher-order topology" in the community. The higher-order topological
  systems in our work are bulk topological phases with necessary crystalline
  symmetry protections and thus admit a higher-order bulk-boundary
  correspondence}\BibitemShut {NoStop}%
\end{thebibliography}

\end{document}